\documentclass[a4paper,11pt]{article}
\pdfoutput=1 % if your are submitting a pdflatex (i.e. if you have
\usepackage{jheppub} % for details on the use of the package, please
\usepackage[T1]{fontenc} % if needed

\usepackage{braket}
\usepackage{amsmath}
\usepackage{slashed}
\usepackage[capitalize]{cleveref}

\def\Lst{{\Lambda^*}}
\def\Lb{{\Lambda_b}}
\def\C#1{{\cal C}_{#1}}
\newcommand{\abs}[1]{\left\lvert#1\right\rvert}

\preprint{LPT-Orsay-19-08}

\title{\boldmath Angular analysis of the rare decay $\Lb\to \Lambda(1520)(\to N\bar{K})\ell^+\ell^-$}

\author{S. Descotes-Genon,}

\author{M. Novoa-Brunet}

\affiliation{Laboratoire de Physique Th\'eorique (UMR 8627),\\
CNRS, Univ. Paris-Sud, Universit\'e Paris-Saclay, 91405 Orsay Cedex, France}

\emailAdd{sebastien.descotes-genon@ijclab.in2p3.fr}
\emailAdd{martin.novoa@ijclab.in2p3.fr}

\abstract{We study the differential decay rate for the rare decay $\Lb\to \Lambda(1520)(\to N\bar{K})\ell^+\ell^-$ where $\ell$ is a light lepton and $N\bar{K}=pK^-,n\bar{K}^0$, as  this decay mode can provide new and complementary constraints on the Wilson coefficients in $b\to s\ell^+\ell^-$ transitions compared to other modes. We provide a determination of the complete angular distribution, assuming unpolarised $\Lb$ baryons and neglecting the lepton mass. The resulting angular observables are expressed in terms of helicity amplitudes involving hadronic form factors within the Standard Model and New Physics models with chirality-flipped operators. We study these observables at low and large $\Lambda$ recoils, using effective theories to determine relations among the hadronic form factors involved. As there is currently no determination of the form factors available from lattice simulations or light-cone sum rules, we perform a first illustration of the sensitivity of some observables to New Physics contributions using hadronic inputs from quark models.
}

\begin{document}
\maketitle
\flushbottom

\section{Introduction}

The flavour-changing neutral current transition $b\to s\ell^+\ell^-$ has raised considerable interest over the last years in connection with searches for physics beyond the Standard Mode (SM). Indeed, this transition is CKM and loop suppressed in the SM and therefore very sensitive to New Physics (NP). Many processes involving $b\to s\mu^+\mu^-$ at the quark level have been measured by several experiments, showing a series of deviations from the SM in the branching ratios for $B\to K\mu^+\mu^-$~\cite{Aaij:2014pli}, $B\to K^*\mu^+\mu^-$~\cite{Aaij:2014pli,Aaij:2013iag,Aaij:2016flj}, $B_s\to \phi\mu^+\mu^-$~\cite{Aaij:2015esa} as well as for the optimised angular observables~\cite{Matias:2012xw,Descotes-Genon:2013vna} in $B\to K^*\mu^+\mu^-$~\cite{Aaij:2013aln,Aaij:2015oid,Abdesselam:2016llu,ATLAS:2017dlm,CMS:2017ivg}.
Moreover, the comparison of $b\to s\mu^+\mu^-$ and $b\to se^+e^-$ through the measurements of $R_K$~\cite{Aaij:2014ora}, $R_{K^*}$~\cite{Aaij:2017vbb} and $B\to K^*\ell^+\ell^-$ angular observables~\cite{Aaij:2015dea,Wehle:2016yoi} for several values of the dilepton invariant mass hint at a violation of lepton flavour universality (LFU).

These deviations can be explained in a consistent way within a model-independent effective approach: it requires NP contributions to the short-distance Wilson coefficients associated with a limited set of operators describing $b\to s\mu^+\mu^-$ transitions. A recent combined analysis of these observables~\cite{Capdevila:2017bsm} indeed singles out some NP scenarios preferred over the SM with a significance at the $5\,\sigma$ level~{\footnote{This confirms the scenarios already highlighted in earlier analyses, mainly restricted to $b\to s\mu^+\mu^-$ processes~\cite{Descotes-Genon:2013wba,Descotes-Genon:2015uva,Altmannshofer:2015sma,Hurth:2016fbr}.}.} The significance for these NP scenarios considering the LFU-violating observables $R_K$ and $R_{K^*}$ but excluding $b\to s\mu^+\mu^-$ observables is at the 3-$4\,\sigma$ level~\cite{Altmannshofer:2017yso,DAmico:2017mtc,Geng:2017svp,Ciuchini:2017mik,Hiller:2017bzc,Hurth:2017hxg,Arbey:2018ics}. The corresponding violation of LFU between muons and electrons is indeed significant, around $25\%$ of the SM value for the semileptonic operator $O_{9\mu}$, with several scenarios showing an equivalent ability to explain the observed deviations.

In this context, it is particularly interesting to confirm and constrain further the scenarios of New Physics in $b\to s\ell^+\ell^-$ transitions. On the experimental side, one can increase the size of the data sample (as currently done by LHCb, CMS and ATLAS), add new observables and exploit different experimental environments (soon provided by Belle II)~\cite{Alguero:2019pjc}. On the theory side, one can improve the determination of hadronic contributions to these decays, i.e. (local) form factors and (non-local) charm-loop contributions~\cite{Bobeth:2017vxj}. It is also interesting to consider other hadronic decays corresponding to the same quark-level transition. Indeed LHCb provides information on decays not only of mesons but also of baryons containing a quark $b$.  The theoretical analysis of these decays does not stand at the same level as for meson decays (in particular for the determination of the form factors and the estimation of charm-loop contributions) but it can provide interesting cross checks of the deviations already observed.

For instance, the $\Lb\to \Lambda(1116)(\to N\pi)\mu^+\mu^-$ has been investigated theoretically in Refs.~\cite{Gutsche:2013pp,Boer:2014kda,Roy:2017dum, Das:2018iap,Das:2018sms,Blake:2017une} and measured by the LHCb collaboration in Refs.~\cite{Aaij:2015xza,Aaij:2018gwm} for both the differential decay rate and the angular observables. There seems to be a trend for the branching ratio to be lower than the SM expectations at large-$\Lambda$ recoil and larger at low-$\Lambda$ recoil (although compatible within errors), whereas the measured angular asymmetries at low-$\Lambda$ recoil did not indicate any deviation from the SM expectations. A study~\cite{Meinel:2016grj} based on the branching ratio data at low-$\Lambda$ recoil using Bayesian statistics and lattice inputs for the $\Lb\to\Lambda(1116)$ form factors~\cite{Detmold:2016pkz} favoured positive shifts in $\C{9}$ (with an opposite sign with respect to the global fits mentioned earlier), but the exploitation of only a subset of the available data, the large experimental uncertainties and the lack of knowledge about non-local contributions for this decay make the interpretation of this result delicate.

It is thus worth testing the $b\to s\ell^+\ell^-$ transition further in the baryon sector. Interestingly, the LHCb search for pentaquarks states in $\Lb\to pK^- J/\psi$ provides information on $\Lb\to \Lambda(\to pK^-)\ell^+\ell^-$ for a dilepton invariant mass $q^2$ around the $J/\psi$ mass, where $\Lambda$ is any intermediate baryon with the appropriate quantum numbers. As indicated in Fig.~3 of Ref.~\cite{Aaij:2015tga}, the dominant contribution comes from $\Lambda(1520)$ ($J^P=3/2^-$). For a $Kp$ invariant mass around 1.5 GeV, there is a  contamination coming from two other states, $\Lambda(1405)$ (with a mass below the $N\bar{K}$ threshold, but sufficiently wide to provide a contribution to this decay) and $\Lambda(1600)$. Following the LHCb analysis, these two states contribute at similar levels and they might be discriminated from $\Lambda(1520)$ thanks to their different spin and parity ($J^P=1/2^\pm$ rather than $3/2^-$) -- for instance, this could be implemented through an angular analysis, although this demanding approach would require a significant number of events. This dominance of $\Lambda(1520)$ for a $Kp$ invariant mass around 1.5 GeV, which has been observed for $q^2=m_{J/\psi}^2$, may hold for other values of the dilepton invariant mass. For instance, the $\Lb\to pK^-\gamma$ decay has been investigated to determine its potential in determining the polarisation of the photon in the $b\to s\gamma$ transition using polarised $\Lb$ baryons~\cite{Legger:2006cq,Hiller:2007ur}. These studies involve models for the $pK^-$ invariant mass spectrum where $\Lambda(1520)$ is again prominent, but this time for $q^2=0$ (see Fig.~1 in Ref.~\cite{Legger:2006cq}).
One may thus hope that for a large range dilepton invariant mass $q^2$, the contribution from  $\Lambda(1520)$ remains large
and could be extracted from the signal observed in $\Lb\to pK^-\mu^+\mu^-$~\cite{Aaij:2017mib}, so that the decay $\Lb\to \Lambda(1520)\ (\to N K)\ell^+\ell^-$ should be accessible and could be studied in detail at LHCb, both for the branching ratio and for the angular observables.

Compared to $\Lambda(1116)$, a decay involving $\Lambda(1520)$ (from now on denoted as $\Lst$) feature two differences already discussed: the spin of the intermediate $\Lst$ state is higher ($J^P=3/2^-$ rather than $1/2^+$) and it decays into $pK^-$ under the strong interaction (rather than into $p\pi$ under the weak interaction). Obviously the same issues exist in both cases concerning the uncertainties on hadronic contributions~\cite{Capdevila:2017ert}: preliminary lattice determinations of the form factors have been presented in Ref.~\cite{Meinel:2016cxo} and the question of non-local contributions could be understood based on data-driven methods similar to Refs.~\cite{Bobeth:2017vxj,Blake:2017fyh}, involving light-cone sum rules similar to Refs.~\cite{Feldmann:2011xf,Khodjamirian:2010vf,Khodjamirian:2012rm,Wang:2015ndk} It is thus already interesting to discuss the general structure of this decay and the observables that can be obtained, even before these issues are completely resolved.

The article is organised in the following way. We discuss the effective Hamiltonian for $b\to s\ell^+\ell^-$ transitions and the kinematics of  $\Lb\to \Lst(\to N\bar{K})\ell^+\ell^-$ with $N\bar{K}=pK^-,n\bar{K}^0$ in Sec.~\ref{sec:effhamkin}. We consider different hadronic inputs for this transition in Sec.~\ref{sec:hadronic}, namely the $\Lb\to \Lst$ form factors and the description of the $\Lst\to N\bar{K}$ decay. We compute the helicity amplitudes and the angular observables  and discuss phenomenology aspects of this decay in Sec.~\ref{sec:pheno}, before drawing a few conclusions in Sec.~\ref{sec:concl}. Technical considerations concerning the kinematics and the free solutions in several rest frames as well as cross checks of our results with earlier work are collected in appendices.

\section{General framework}\label{sec:effhamkin}

\subsection{Effective Hamiltonian}

It is possible to analyse $b\to s\ell^+\ell^-$ decays using a model-independent approach, namely the effective Hamiltonian~\cite{Grinstein:1987vj,Buchalla:1995vs} where heavy degrees of freedom have been integrated out in short-distance Wilson coefficients $\C{i}$, leaving only a set of operators $O_i$ describing the physics on long distances
\begin{equation}\label{eq:hameff}
{\cal H}_{\rm eff}=-\frac{4G_F}{\sqrt{2}} V_{tb}V_{ts}^*\sum_i \C{i}  O_i,
\end{equation}
(up to small corrections proportional to $V_{ub}V_{us}^*$ in the SM). The factorisation scale for the Wilson coefficients is chosen to be $\mu_b=$ 4.8 GeV. We focus  our attention on the operators
\begin{align}
{\mathcal{O}}_{7} &= \frac{e}{16 \pi^2} m_b
(\bar{s} \sigma_{\mu \nu} P_R b) F^{\mu \nu} ,&
{\mathcal{O}}_{{7}^\prime} &= \frac{e}{16 \pi^2} m_b
(\bar{s} \sigma_{\mu \nu} P_L b) F^{\mu \nu} , \nonumber
%\label{O7}
\\
{\mathcal{O}}_{9\ell} &= \frac{e^2}{16 \pi^2}
(\bar{s} \gamma_{\mu} P_L b)(\bar{\ell} \gamma^\mu \ell) ,&
{\mathcal{O}}_{{9}^\prime\ell} &= \frac{e^2}{16 \pi^2}
(\bar{s} \gamma_{\mu} P_R b)(\bar{\ell} \gamma^\mu \ell) , \nonumber
\\
\label{eq:O10}
{\mathcal{O}}_{10\ell} &=\frac{e^2}{16 \pi^2}
(\bar{s}  \gamma_{\mu} P_L b)(  \bar{\ell} \gamma^\mu \gamma_5 \ell) ,&
{\mathcal{O}}_{{10}^\prime\ell} &=\frac{e^2}{16\pi^2}
(\bar{s}  \gamma_{\mu} P_R b)(  \bar{\ell} \gamma^\mu \gamma_5 \ell) ,
\end{align}
where $P_{L,R}=(1 \mp \gamma_5)/2$ and $m_b \equiv m_b(\mu_b)$ denotes the running $b$ quark mass in the $\overline{\mathrm{MS}}$ scheme.
In the SM, three operators play a leading role in the discussion, namely the electromagnetic operator $O_7$ and the semileptonic operators $O_{9\ell}$ and $O_{10\ell}$, differing with respect to the chirality of the emitted charged leptons. NP contributions could either modify the value of the short-distance Wilson coefficients $\C{7,9,10}$, or make other operators contribute in a significant manner, such as the chirality-flipped operators $O_{7',9',10'}$ defined above, or other operators (scalar, pseudoscalar, tensor). We will focus on the effect of SM operators and their chirality-flipped counterparts, although we will discuss the impact of the other SM operators (four-quark operators $O_{1-6}$ and $O_{8g}$)  briefly in Sec.~\ref{sec:pheno}.

\subsection{Kinematics}\label{sec:kinematics}

\begin{figure}
    \centering
    \includegraphics{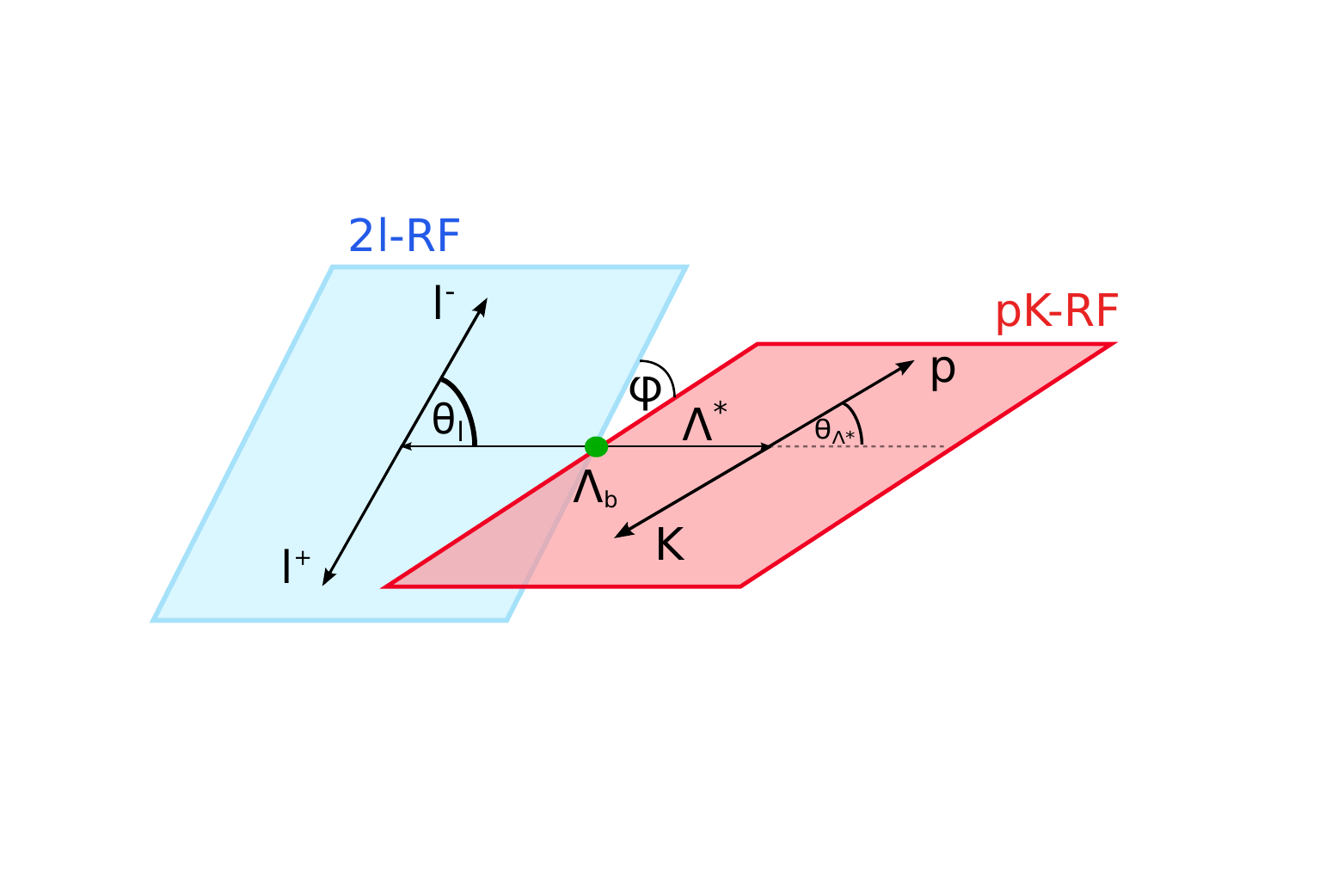}
    \caption{Kinematics of the four-body $\Lb\to\Lst(\to N\bar{K})\ell^+\ell^-$ decay. The angles are defined in the corresponding rest frames, indicated in colours.}
    \label{fig:kinem}
\end{figure}

We consider the decay chain with the corresponding momenta for the various particles and their spin projections along the $z$-axis of the rest frame of the decaying particle
\begin{eqnarray}
\Lb(p,s_{\Lb}) &\to & \Lst(k,s_\Lst) \ell^+(q_1)\ell^-(q_2), \\
\Lst(k,s_\Lst)&\to & N(k_1,s_N) K(k_2)  \qquad (N\bar{K}=pK^-,n\bar{K}^0),
\end{eqnarray}
where we denote $\Lst(1520)$ as $\Lst$ and we have
\begin{equation}
q^\mu=q_1^\mu+q_2^\mu, \qquad k^\mu=k_1^\mu+k_2^\mu, \qquad p^\mu=q^\mu+k^\mu.
\end{equation}
We can introduce the same kinematics as for semileptonic four-body $B$-meson decays, leading to four independent variables chosen as the dilepton invariant mass $q^2$, the angles $\theta_\Lst$ and $\theta_\ell$ with respect to the $z$ axis and the angle between the hadronic and leptonic planes $\phi$, following the same LHCb conventions as for $\Lb\to\Lambda(\to N\pi)\ell^+\ell^-$~\cite{ Blake:2017une,Aaij:2015xza,Aaij:2018gwm} (up to the identifications $\theta_\Lst=\theta_b$ and $\phi=\chi$)
recalled in Fig.~\ref{fig:kinem}. The CP-conjugate mode can also be described using the same formalism, where an appropriate redefinition of the angle ensures that the angular observables will have the same form, up to the complex conjugation of the weak phases~\cite{Blake:2017une,Gratrex:2015hna}.

The differential decay rate can be written
\begin{equation}
d\Gamma=\frac{|{\mathcal{M}}|^2}{2m_\Lb}d\Phi_4(p;k_1,k_2,q_1,q_2),
\end{equation}
where the phase space
\begin{equation}
d\Phi_n(P;p_1,\ldots, p_n)=(2\pi)^4\ \delta^{(4)}(P-\sum_i p_i)\ \prod_i \frac{d^3 p_i}{(2\pi)^3 2E_i},
\end{equation}
can be decomposed iteratively as~\cite{Patrignani:2016xqp}
\begin{equation}
\int d\Phi_4 \frac{|{\mathcal{M}}|^2}{2m_\Lb}=\int \frac{dq^2}{2\pi}\frac{dk^2}{2\pi}d\Phi_2(k;k_1,k_2)d\Phi_2(q;q_1,q_2)d\Phi_2(p;q,k) \frac{|{\mathcal{M}}|^2}{2m_\Lb},
\end{equation}
with the two-body phase space of the form
\begin{equation}
\int d\Phi_2(k;k_1,k_2) X(k;k_1,k_2)=\int \frac{d\Omega_k}{4\pi} \frac{1}{8\pi}\frac{\sqrt{\lambda(k^2,k_1^2,k_2^2)}}{k^2}
X(k;k_1,k_2),
\end{equation}
with $d\Omega_k$ the element of integration of the the solid angle for $k$ and the  K\"all\'en function is
\begin{equation}
\lambda(k^2,k_1^2,k_1^2)=k^4+k_1^4+k_2^4-2k^2k_1^2-2k^2k_2^2-2k_1^2k_2^2.
\end{equation}
The re-expression of the two-body phase space differential elements yields
\begin{equation}\label{eq:phasespace}
\int d\Phi_4 \frac{|{\mathcal{M}}|^2}{2m_\Lb}
=\frac{1}{(2\pi)^2(8\pi)^4}\int dk^2\ (\beta_\Lb\beta_{N\bar{K}}\beta_\ell)
\ (dq^2\ d\cos\theta_\ell\ d\cos\theta_\Lst\ d\phi) \frac{|{\mathcal{M}}|^2}{2m_\Lb},
\end{equation}
with
\begin{equation}
\beta_\Lb=\frac{\sqrt{\lambda(m_{\Lb}^2,k^2,q^2)}}{m_{\Lb}^2},\qquad \beta_{N\bar{K}}=\frac{\sqrt{\lambda(k^2,m_N^2,m_{\bar{K}}^2)}}{k^2},\qquad \beta_\ell=\sqrt{1-\frac{4m_\ell^2}{q^2}}.
\end{equation}

\subsection{Helicity amplitudes}

It is well known that such a decay chain is best analysed by performing a decomposition according to helicity amplitudes, as discussed in Ref.~\cite{Haber:1994pe} and illustrated for $b\to s\ell^+\ell^-$ decays in Ref.~\cite{Gratrex:2015hna}. In particular, it proves very useful to introduce a vector basis which can be seen as the polarisation of an intermediate virtual boson decaying into the dilepton pair~\footnote{This interpretation is discussed in detail in Ref.~\cite{Gratrex:2015hna}, where it is shown to be valid in full generality in the absence of tensor operators in the effective Hamiltonian.}, defined in the dilepton rest frame as
\begin{equation}\label{eq:polarisations}
\varepsilon^\mu(0)=\left(\begin{array}{c}0\\0\\0\\1\end{array}\right), \quad
\varepsilon^\mu(+)=\frac{1}{\sqrt{2}}\left(\begin{array}{c}0\\-1\\-i\\0\end{array}\right), \quad
\varepsilon^\mu(-)=\frac{1}{\sqrt{2}}\left(\begin{array}{c}0\\1\\-i\\0\end{array}\right), \quad
\varepsilon^\mu(t)=\left(\begin{array}{c}1\\0\\0\\0\end{array}\right).
\end{equation}
in agreement with Ref.~\cite{Haber:1994pe}. Boosts can be used to define this basis in other reference frames.
We can also easily define the scalar and time-like vectors in a general way as
\begin{equation}\label{eq:polarisations_general}
\varepsilon^\mu(0) = \frac{e^\mu}{\sqrt{|e^2|}},
\qquad \varepsilon^\mu(t) =\frac{q^\mu}{\sqrt{q^2}},
\qquad e^\mu = p^\mu+k^\mu-\frac{q^\mu}{q^2}(m_{\Lb}^2-m_\Lst^2),
\end{equation}
with $e^2=-\lambda(m_\Lb^2,m_\Lst^2,q^2)/q^2$.

We have the completeness and orthogonality relations for $\lambda,\lambda'=t,0,+,-$
\begin{equation}
\varepsilon^{*\mu}(\lambda) \varepsilon_{\mu}(\lambda') = g_{\lambda\lambda'}, \qquad
\sum_{\lambda,\lambda'=t,0,+,-} \varepsilon^{*\mu}(\lambda) \varepsilon^\nu(\lambda') g_{\lambda\lambda'} = g^{\mu\nu},
\end{equation}
where $g$ is defined as diagonal in the polarisation space, with $g_{tt}=-g_{00}=-g_{++}=-g_{--}=1$. Assuming factorisation between the hadronic and the leptonic parts, and focusing on operators with a single Lorentz index, we have  for the $\Lambda\to \Lst\ell^+\ell^-$ part of this decay chain
\begin{equation}
\begin{aligned}\label{eq:LbtoLstll}
\langle \Lst \ell^+\ell^-|O^H_\mu O^{L;\mu} |\Lb\rangle
 =&\langle \Lst |O^H_\mu |\Lb\rangle   g^{\mu\nu}  \langle \ell^+\ell^-|O^L_\nu |0\rangle\\
 =&\sum_{\lambda=t,0,+,-}g_{\lambda\lambda'} (\varepsilon^{*\mu}(\lambda)\langle \Lst |O^H_\mu |\Lb\rangle) (\varepsilon^\nu(\lambda)  \langle \ell^+\ell^-|O^L_\nu |0\rangle)\\
  =&\sum_{\lambda=t,0,+,-}g_{\lambda\lambda'} \sum_i  f^i_\lambda
\ (\bar{u}_\Lst^\alpha \varepsilon^{*\mu}(\lambda)\Gamma^{Hi}_{\mu\alpha} u_\Lb)
\ (\bar{u}_{\ell^-} \varepsilon^\nu(\lambda)\Gamma^{L}_\nu v_{\ell^-}),
 \end{aligned}
\end{equation}
which defines the hadronic and leptonic helicity amplitudes of interest, where we express the results in terms of the solutions for free fermions $u$ and $v$ (see App.~\ref{app:kinematics} for explicit expressions) and the helicity form factors $f^i_\lambda$ to be defined in more detail in section \ref{sec:hadronic}.

\subsection{Propagation and decay of $\Lst$}\label{sec:lambdastar}

Once $\Lambda\to\Lst\ell^+\ell^-$ has been described, we still have to include the propagation and the decay of the $\Lst$ baryon, which has the quantum numbers $J^{P}=3/2^-$. This is usually done in the Rarita-Schwinger framework~\cite{Rarita:1941mf}, with a field $\psi^\alpha_a$ combining a spinor index $a$ and a vector index $\alpha$
(in the following, the spinor index $a$ will often be kept implicit for simplicity). The corresponding free Lagrangian reads
\begin{equation}\label{eq:freeRS}
{\cal L}=\bar\psi_\mu \frac{i}{2}\{ \sigma^{\mu\nu},(i\slashed{\partial}-m_\Lst)\} \psi_\nu.
\end{equation}
The solutions $u_a^\alpha(k,s_\Lst)$ obey then the following properties:
\begin{equation}
\slashed{k}u^{\alpha}=m_\Lst u^{\alpha},\qquad \gamma_\alpha u^{\alpha}=0,\qquad k_\alpha u^{\alpha}=0.
\end{equation}
Following Ref.~\cite{Huang:2003ym}, one can determine the explicit solutions for $s_\Lst=-3/2,-1/2,1/2,3/2$ as
\begin{equation}\label{eq:sol32}
u_a^\alpha(k,s_\Lst)=\sum_{\lambda=-1,0,1}\sum_{r=-1/2,1/2}
     \delta_{\lambda+r,s_\Lst}
      \left\langle 1,\lambda;\frac{1}{2},r\bigg| 1,\frac{1}{2},\frac{3}{2},s_\Lst\right\rangle  \varepsilon^\alpha_\lambda(k)u^r_a(k),
\end{equation}
and a similar construction can be found for the anti-fermion solutions. This is used in App.~\ref{app:kinematics} to determine the appropriate solutions of the $\Lst$ equation of motion in both $\Lb$ and $\Lst$ rest frames.

The free propagator derived from Eq.~(\ref{eq:freeRS}) reads
\begin{equation}\label{eq:propagatorRS}
G^{\mu\nu}=\frac{i(\slashed{k}+m_\Lst)\Delta^{\mu\nu}}{k^2-m_\Lst^2},
\qquad \Delta^{\mu\nu}=g^{\mu\nu}{\rm Id}-\frac{1}{3}\gamma^\mu\gamma^\nu-\frac{2k^\mu k^\nu}{3m_\Lst^2}-\frac{\gamma^\mu k^\nu-\gamma^\nu k^\mu}{3m_\Lst},
\end{equation}
where ${\rm Id}$ denotes the identity matrix in the Dirac matrix space.
We checked that the summation formula expected from general unitarity arguments (see for instance Ref.~\cite{Schwartz:2013pla})
\begin{equation}
\sum_{s_\Lst=-3/2}^{3/2} u^\mu_a(k,s_\Lst) \bar{u}^\nu_b(k,s_\Lst)
  = -(\slashed{k}+m_\Lst)\Delta^{\mu\nu}_{ab},
\end{equation}
is indeed satisfied by the solution Eq.~(\ref{eq:sol32}) given in App.~\ref{app:kinematics}.

One should note that the tensor $\Delta^{\mu\nu}$ involved in the propagator is not the projector on the spin-3/2 component given by
\begin{equation}\label{eq:projectorRS}
P^{\mu\nu}=g^{\mu\nu}-\frac{1}{3}\gamma^\mu\gamma^\nu-\frac{1}{3m_\Lst^2}(\slashed{k}\gamma^\mu k^\nu+k^\mu\gamma^\nu \slashed{k}).
\end{equation}
By construction, a field of the form $\psi^\alpha_a$ contains both spin-1/2 and spin-3/2 components: only
4 components of the 16-component field $\psi^\alpha_a$ are actually needed to describe the spin-3/2 part.
Even though the Rarita-Schwinger construction aims at describing a spin-3/2 object only, it turns out that one cannot modify the kernel in Eq.~(\ref{eq:freeRS}) to keep only on the spin-3/2 component of the field, as the projected kernel cannot be inverted~\cite{Benmerrouche:1989uc}. Attempts to enforce the projection at the level of the propagator led to theories with unwanted properties, such as spurious poles affecting higher orders in perturbation theory~\cite{Johnson:1960vt,Velo:1970ur,Kobayashi:1987rt}. In practice, the quantisation must be performed with the whole field $\psi^\alpha_a$, in the presence of constraints that will ensure that only the spin-3/2 component of the field is actually physical~\cite{Pascalutsa:1998pw}.

The problem is even made more acute in the case of an interacting spin-$3/2$ theory, as the interaction term should be compatible with the quantisation of the theory. This led to a significant amount of debate concerning the description of the $\pi N\Delta$ interaction at low energies, which can be used as a template to describe the $K N \Lst $ interaction (the quantum numbers are the same, apart from the opposite parity of the $\Delta$ and $\Lst$ fermions). A first type of effective interaction at low energies was proposed (here translated to the $K N \Lst $ case)~\cite{Nath:1971wp}
\begin{equation}\label{eq:L1RS}
{\cal L}_1=g m_\Lst \bar\psi_\mu(g^{\mu\nu}+a\gamma^\mu\gamma^\nu)\gamma_5\Psi\partial_\nu\phi + h.c.,
\end{equation}
where $\Psi$ denotes the spin-$1/2$ $N$ field and $\phi$ the spin-0 $K$ field (coming with a derivative due to the pseudo-Goldstone nature of the kaon), and $a$ is an off-shell parameter that is relevant only for loop computations.

As discussed in Ref.~\cite{Pascalutsa:1998pw}, this interaction is simple, but once used to build an interacting theory, it involves not only the physical spin-$3/2$ components of the Rarita-Schwinger field $\psi_\mu$, but also unphysical spin-1/2 components, leading to problems of causality and to a significant contribution from spin-$1/2$ background underneath the $\Lst$ (or $\Delta$) resonance. In Ref.~\cite{Pascalutsa:1998pw}, an alternative interaction has been proposed
\begin{equation}\label{eq:L2RS}
{\cal L}_2=g\varepsilon^{\mu\nu\alpha\beta}(\partial_\mu\bar\psi_\nu)\gamma_\alpha \Psi\partial_\beta\phi + h.c.
\end{equation}
This choice is suggested by the invariance of the free massless theory under gauge transformations of $\psi_\mu$:  it is compatible with the quantisation of the theory under constraints, and using the projector Eq.~(\ref{eq:projectorRS}) and the propagator Eq.~(\ref{eq:propagatorRS}), it can be shown easily that it involves only the spin-3/2 part of $\psi^\alpha$.

Fortunately, we do not have to take sides on this issue here. Indeed,
these two choices of interaction term will yield actually the same result for the branching ratio of interest here. This is in agreement with the fact that we use these interactions only for a tree-level interaction with on-shell particles, and it will provide a further cross-check of our results.

\subsection{Narrow-width approximation}

The $\Lst$ propagation and decay can thus be included in the description Eq.~(\ref{eq:LbtoLstll}) as
\begin{equation}
\begin{aligned}\label{eq:M}
{\cal M}=&\langle (N\bar{K})_\Lst \ell^+\ell^-|O^H_\mu O^{L;\mu} |\Lb\rangle\\
  =&i \sum_{\lambda=t,0,+,-} \sum_i  \sum_{s_\Lst}
\bar{u}_N G u_\Lst
\
\frac{i}{k^2-m_\Lst^2}
\ (\bar{u}_\Lst^\alpha (\varepsilon_\lambda^{*\mu}\Gamma^{Hi}_{\mu\alpha}) u_\Lb)\  f^i_\lambda
\ (\bar{u}_{\ell^-} (\varepsilon_{\lambda}^\nu\Gamma^{L}_\nu) v_{\ell^-}),
 \end{aligned}
\end{equation}
where $G$ is a momentum-dependent quantity defined as $\langle N K | {\cal L}_i |\Lst \rangle = \bar{u}_N G(k_1,k_2) u_\Lst$  from the interaction Lagrangians Eqs.~(\ref{eq:L1RS}) or (\ref{eq:L2RS}). The decay rate can be computed as
\begin{equation}
\int d\Gamma=\int d\Phi_4 \frac{|\overline{\mathcal{M}}|^2}{2m_\Lb},
\qquad |\overline{\mathcal{M}}|^2=\frac{1}{2}\sum_{s_\Lb} \sum_{s_N} |{\mathcal{M}}|^2,
\end{equation}
where we summed over the final spins and averaged over the initial spins, assuming that the $\Lb$ baryon is produced essentially in an unpolarised way at the LHC~\cite{Aaij:2013oxa,Sirunyan:2018bfd}.

Following Ref.~\cite{Altmannshofer:2008dz},
we modify the propagator of the $\Lst$ baryon to take into account the width of the resonance, but treat it as narrow ($\Gamma_\Lst \ll m_\Lst$)
\begin{equation}
\begin{aligned}
\int  d\Phi_4 \frac{|\overline{\mathcal{M}}|^2}{2m_\Lb}
 &= \int d\tilde\Phi\ dk^2 \frac{|\overline{\cal N}|^2}{(k^2-m^2)^2}\\
 &\to \int d\tilde\Phi\ dk^2 \frac{|\overline{\cal N}|^2}{(k^2-m^2)^2+(m_\Lst \Gamma_\Lst)^2}\\
 &\to \int d\tilde\Phi\ dk^2 |\overline{\cal N}|^2 \ \frac{\pi}{m_\Lst \Gamma_\Lst} \delta(k^2-m_\Lst^2)
 = \int d\tilde\Phi  |\overline{\cal N}|^2_{k^2=m_\Lst^2} \ \frac{\pi}{m_\Lst \Gamma_\Lst},
 \end{aligned}
 \end{equation}
where $d\tilde\Phi$ describes the phase space without the integration with respect to $dk^2$ as shown in Eq.~(\ref{eq:phasespace}), and $\cal N$ is defined from the matrix element $\cal M$ in Eq.~(\ref{eq:M}) as
\begin{equation}
    {\cal N}=(k^2-m_\Lst^2){\cal M}.
\end{equation}
Up to a phase space, the branching ratio is the product of three matrix elements corresponding to the helicity amplitude for the leptonic part, the helicity amplitude for the $\Lb\to \Lst$ hadronic part and the matrix element for the $\Lst \to N\bar{K}$ decay. We finally obtain
\begin{equation}\label{eq:dGamma}
\int d\Gamma
=\left. \int dq^2d\cos\theta_\ell d\cos\theta_\Lst d\phi \frac{1}{2^{15}\pi^5m_\Lb m_\Lst\Gamma_\Lst} (\beta_\Lb\beta_{N\bar{K}}\beta_\ell)
 |\overline{\mathcal{N}}|^2  \right|_{k^2=m_\Lst^2}.
\end{equation}

\section{Hadronic matrix elements}\label{sec:hadronic}

Since the general framework of the kinematics and helicity amplitudes has been set up, we can turn to the description of the hadronic part of the decay through form factors.

\subsection{$\Lb\to\Lst$ vector form factors}

The hadronic matrix elements can be decomposed using the spinors for $\Lb$ and $\Lst$, and inserting all the possible Dirac structures taking into account the parity and the e.o.m constraints. In the case of the vector form/axial operators, there are four structures, and thus four form factors, once the equations of motion for $\Lb$ and $\Lst$ are taken into account~\cite{Mott:2011cx}.
As seen before, they are better defined using helicity form factors~\cite{Feldmann:2011xf,Boer:2014kda}, which corresponds to choosing combinations of these Dirac matrices so that they are orthogonal to the polarisation eigenvectors defined in Eq.~(\ref{eq:polarisations}). This leads to the following definition
\begin{equation}
\begin{aligned}\label{eq:ffVA}
\braket{\Lst | \bar s \gamma^\mu b|\Lb}=
& \bar u_\alpha(k,s_\Lst)\biggl\{p^\alpha\biggl[f_t^V(q^2) (m_{\Lb}-m_\Lst)\frac{q^\mu}{q^2}\\
&+ f_0^V(q^2) \frac{m_{\Lb}+m_\Lst}{s_+}(p^\mu +k^\mu-\frac{q^\mu}{q^2}(m_{\Lb}^2-m_\Lst^2))\\
&+ f_\perp^V(q^2)(\gamma^\mu-2\frac{m_\Lst}{s_+}p^\mu -2\frac{m_{\Lb}}{s_+}k^\mu)\biggr]\\
&+ f_g^V(q^2) \left[g^{\alpha\mu}+m_\Lst\frac{p^\alpha}{s_-} \left(\gamma^\mu - 2 \frac{k^\mu}{m_\Lst} +2 \frac{m_\Lst p^\mu +m_\Lb k^\mu}{s_+}\right)\right]\biggr\}u(p,s_{\Lb}),\\
\braket{\Lst | \bar s \gamma^\mu \gamma ^5 b|\Lb}=& -\bar u_\alpha(k,s_\Lst)\gamma^5\biggl\{p^\alpha\biggl[
f_t^A(q^2) (m_{\Lb}+m_\Lst)\frac{q^\mu}{q^2}\\
    &+ f_0^A(q^2) \frac{m_{\Lb}-m_\Lst}{s_-}(p^\mu +k^\mu-\frac{q^\mu}{q^2}(m_{\Lb}^2-m_\Lst^2))\\
&+ f_\perp^A(q^2)(\gamma^\mu+2\frac{m_\Lst}{s_-}p^\mu -2\frac{m_{\Lb}}{s_-}k^\mu)\biggr]\\
&+ f_g^A(q^2) \left[g^{\alpha\mu}-m_\Lst\frac{p^\alpha}{s_+} \left(\gamma^\mu + 2 \frac{k^\mu}{m_\Lst} -2 \frac{m_\Lst p^\mu -m_\Lb k^\mu}{s_-}\right)\right]\biggr\}u(p,s_{\Lb}).
\end{aligned}
\end{equation}
We have introduced
\begin{equation}
s_\pm=(m_{\Lb}\pm m_\Lst)^2-q^2.
\end{equation}
 We have used similar normalisations to the $\Lb\to\Lambda(1116)$ form factors chosen in Refs.~\cite{Feldmann:2011xf,Boer:2014kda} for $f_t,f_0,f_\perp$ so that in the limit where the three form factors are set to 1, one recovers a point-like behaviour
 $\bar u_\alpha(k,s_\Lst) p^\alpha\gamma^\mu(\gamma_5) u(p,s_\Lb)$. However, in the $\Lst$ case, a fourth form factor, $f_g$, arises~\cite{Mott:2011cx,Meinel:2016cxo,Boer:2018vpx}.
Further constraints arise by considering the limit $q^2\to0$:
since there are no physical state with $\bar{s}b$ quantum numbers and a a vanishing mass, the matrix elements cannot exhibit any singularity at $q^2=0$, which leads to the constraints in this limit
 \begin{equation}
     f_t^V(q^2)-f_0^V(q^2)=O(q^2),\qquad
     f_t^A(q^2)-f_0^A(q^2)=O(q^2),
 \end{equation}
 \begin{equation}
\begin{aligned}
    f_t^V(q^2)&=O(1),
    &\ f_0^V(q^2)&=O(1),
    &\ f_\perp^V(q^2)&=O(1),
    &\ f_g^V(q^2)&=O(1),\\
    f_t^{A}(q^2)&=O(1),
    &\  f_0^{A}(q^2)&=O(1),
    &\ f_\perp^{A}(q^2)&=O(1),
    &\ f_g^{A}(q^2)&=O(1).
\end{aligned}
\end{equation}
 Some conditions should also obeyed by the form factors for  $q^2=(m_\Lb-m_\Lst)^2$, where additional $s_-$ factors arise from the normalisation of the free Dirac solutions $u$ and $\bar{u}_\alpha$, see Sec.~\ref{app:kinematics}. We finally obtain the following constraints in this limit
\begin{equation}
\begin{aligned}
    f_t^V(q^2)&=O\left(\textstyle\frac{1}{\sqrt s_-}\right),
    &\ f_0^V(q^2)&=O\left(\textstyle\frac{1}{s_-}\right),
    &\ f_\perp^V(q^2)&=O\left(\textstyle\frac{1}{s_-}\right),
    &\ f_g^V(q^2)&=O(1),\\
    f_t^{A}(q^2)&=O\left(\textstyle\frac{1}{s_-}\right),
    &\  f_0^{A}(q^2)&=O\left(\textstyle\frac{1}{\sqrt s_-}\right),
    &\ f_\perp^{A}(q^2)&=O\left(\textstyle\frac{1}{\sqrt s_-}\right),
    &\ f_g^{A}(q^2)&=O\left(\textstyle\frac{1}{\sqrt s_-}\right).
\end{aligned}\label{eq:endpointbehaviourV}
\end{equation}
At both endpoints, the conditions indicated above are sufficient to ensure the absence of unphysical poles in the hadronic matrix elements, but obviously, form factors exhibiting less singular behaviours are also acceptable.

The choice of helicity form factors means that the matrix elements for each polarisation correspond to a very simple expression in terms of form factors for the vector part
\begin{equation}
\begin{aligned}
&H^V_t(s_{\Lb},s_\Lst)\equiv\varepsilon^*_\mu(t)\braket{\Lst(k,s_\Lst) | \bar s\gamma^\mu b|\Lb(p,s_{\Lb})}\\
&\qquad=f_t^V(q^2)\frac{m_{\Lb}-m_\Lst}{\sqrt{q^2}}\bar u_\alpha(k,s_\Lst)p^\alpha u(p,s_{\Lb}),\\
&H^V_0(s_{\Lb},s_\Lst)\equiv\varepsilon^*_\mu(0)\braket{\Lst(k,s_\Lst) | \bar s\gamma^\mu b|\Lb(p,s_{\Lb})}\\
&\qquad=-f_0^V(q^2)\frac{m_{\Lb}+m_{\Lst}}{s_+}\sqrt{\abs{e^2}}
\bar u_\alpha(k,s_\Lst)p^\alpha u(p,s_{\Lb}),\\
&H^V_\pm(s_{\Lb},s_\Lst)\equiv\varepsilon^*_\mu(\pm)\braket{\Lst(k,s_\Lst) | \bar s\gamma^\mu b|\Lb(p,s_{\Lb})}\\
&\qquad=\left(f_\perp^V(q^2)+f_g^V(q^2)\frac{m_\Lst}{s_-}\right) \bar u_\alpha(k,s_\Lst)p^\alpha \slashed{\varepsilon}^*(\pm) u(p,s_{\Lb}) + f_g^V(q^2) \bar u_\alpha(k,s_\Lst)\varepsilon^{*\alpha}(\pm) u(p,s_{\Lb}),
\end{aligned}
\end{equation}
and for the axial part
\begin{equation}
\begin{aligned}
&H^A_t(s_{\Lb},s_\Lst)\equiv\varepsilon^*_\mu(t)\braket{\Lst(k,s_\Lst) | \bar s\gamma^\mu\gamma^5 b|\Lb(p,s_{\Lb})}\\
&\qquad=-f_t^A(q^2)\frac{m_{\Lb}+m_\Lst}{\sqrt{q^2}}\bar u_\alpha(k,s_\Lst)\gamma^5p^\alpha u(p,s_{\Lb}),\\
&H^A_0(s_{\Lb},s_\Lst)\equiv\varepsilon^*_\mu(0)\braket{\Lst(k,s_\Lst) | \bar s\gamma^\mu\gamma^5 b|\Lb(p,s_{\Lb})}\\
&\qquad=f_0^A(q^2)\frac{m_{\Lb}-m_{\Lst}}{s_-}\sqrt{\abs{e^2}}\bar u_\alpha(k,s_\Lst)\gamma^5 p^\alpha u(p,s_{\Lb}),\\
&H^A_\pm(s_{\Lb},s_\Lst)\equiv\varepsilon^*_\mu(\pm)\braket{\Lst(k,s_\Lst) | \bar s\gamma^\mu\gamma^5 b|\Lb(p,s_{\Lb})}\\
&\qquad=\left(f_\perp^A(q^2)-f_g^A(q^2)\frac{m_\Lst}{s_+}\right) \bar u_\alpha(k,s_\Lst)p^\alpha \slashed{\varepsilon}^*(\pm) \gamma^5 u(p,s_{\Lb}) - f_g^A(q^2) \bar u_\alpha(k,s_\Lst)\varepsilon^{*\alpha}(\pm) \gamma^5 u(p,s_{\Lb}),
\end{aligned}
\end{equation}
where $e$ is the vector defined in Eq.~(\ref{eq:polarisations_general}). Using the expression for the spinor matrix elements given in App.~\ref{app:kinematics}, we obtain for the non-vanishing amplitudes in the vector part
\begin{equation}
\begin{aligned}\label{eq:HplusHminus1}
H^V_t(+1/2,+1/2)&=H^V_t(-1/2,-1/2)=f_t^V (q^2)\frac{m_{\Lb}-m_\Lst}{\sqrt{q^2}}\frac{s_+\sqrt{s_-}}{\sqrt{6}m_\Lst},\\
H^V_0(+1/2,+1/2)&=H^V_0(-1/2,-1/2)=-f_0^V
(q^2)\frac{m_\Lb+m_\Lst}{\sqrt{q^2}}\frac{s_-\sqrt{s_+}}{\sqrt{6}m_\Lst},\\
H^V_+(+1/2,-1/2)&=H^V_-(-1/2,+1/2)=-f_{\perp}^V(q^2)\frac{s_-\sqrt{s_+}}{\sqrt{3}m_\Lst},\\
H^V_+(-1/2,-3/2)&=H^V_-(+1/2,+3/2)=f_{g}^V(q^2)\sqrt{s_+},
\end{aligned}
\end{equation}
and for the axial part
\begin{equation}\label{eq:HplusHminus2}
\begin{aligned}
H^A_t(+1/2,+1/2)&=-H^A_t(-1/2,-1/2)=f_t^A (q^2)\frac{m_{\Lb}+m_\Lst}{\sqrt{q^2}}\frac{s_-\sqrt{s_+}}{\sqrt{6}m_\Lst},\\
H^A_0(+1/2,+1/2)&=-H^A_0(-1/2,-1/2)=-f_0^A(q^2)\frac{m_\Lb-m_\Lst}{\sqrt{q^2}}\frac{s_+\sqrt{s_-}}{\sqrt{6}m_\Lst},\\
H^A_+(+1/2,-1/2)&=-H^A_-(-1/2,+1/2)=f_{\perp}^A(q^2)\frac{s_+\sqrt{s_-}}{\sqrt{3}m_\Lst},\\
H^A_+(-1/2,-3/2)&=-H^A_-(+1/2,+3/2)=-f_{g}^A(q^2)\sqrt{s_-}.
\end{aligned}
\end{equation}

\subsection{$\Lb\to\Lst$ tensor form factors}

A similar discussion takes place in the case of the tensor form factors. The relevant matrix elements are the following, once again defined in order to have structures orthogonal to the polarisation vectors Eq.~(\ref{eq:polarisations})
\begin{equation}\label{eq:ffT}
\begin{aligned}
\braket{\Lst | \bar si \sigma^{\mu\nu}q_\nu b|\Lb}=& - \bar u_\alpha(k,s_\Lst)\biggl\{p^\alpha\biggl[f_0^T(q^2) \frac{q^2}{s_+}(p^\mu +k^\mu-\frac{q^\mu}{q^2}(m_{\Lb}^2-m_\Lst^2))\\
&+ f_\perp^T(q^2)(m_{\Lb} + m_\Lst)(\gamma^\mu-2\frac{m_\Lst}{s_+}p^\mu -2\frac{m_{\Lb}}{s_+}k^\mu)\biggr]\\
&+ f_g^T(q^2)\left[g^{\alpha\mu}+m_\Lst\frac{p^\alpha}{s_-} \left(\gamma^\mu - 2 \frac{k^\mu}{m_\Lst} +2 \frac{m_\Lst p^\mu +m_\Lb k^\mu}{s_+}\right)\right]\biggr\}u(p,s_{\Lb}),\\
\braket{\Lst | \bar si \sigma^{\mu\nu}\gamma^5q_\nu b|\Lb}=& - \bar u_\alpha(k,s_\Lst)\gamma^5\biggl\{p^\alpha\biggl[f_0^{T5}(q^2) \frac{q^2}{s_-}(p^\mu +k^\mu-\frac{q^\mu}{q^2}(m_{\Lb}^2-m_\Lst^2))\\
&+ f_\perp^{T5}(q^2)(m_{\Lb} - m_\Lst)(\gamma^\mu+2\frac{m_\Lst}{s_-}p^\mu -2\frac{m_{\Lb}}{s_-}k^\mu)\biggr]\\
&+ f_g^{T5}(q^2)\left[g^{\alpha\mu}-m_\Lst\frac{p^\alpha}{s_+} \left(\gamma^\mu + 2 \frac{k^\mu}{m_\Lst} -2 \frac{m_\Lst p^\mu -m_\Lb k^\mu}{s_-}\right)\right]\biggr\}u(p,s_{\Lb}).
\end{aligned}
\end{equation}
We have again used similar normalisations to the $\Lb\to\Lambda(1116)$ form factors chosen in Refs.~\cite{Feldmann:2011xf,Boer:2014kda} for $f_0$, $f_\perp$, so that in the limit where the two form factors are set to 1, one recovers a point-like behaviour. In the $\Lst$ case, there is again an additional form factor to be taken into account~\cite{Mott:2011cx,Meinel:2016cxo,Boer:2018vpx}.

As in the vector/axial case, the matrix elements cannot exhibit
a singularity at $q^2=0$ nor $q^2=(m_\Lb-m_\Lst)^2$, which yields the following
constraints for $q^2\to 0$ (see App.~\ref{app:crosscheckgamma} for further detail)
\begin{equation} \label{eq:tensorq2zero}
\begin{aligned}
    f_\perp^T(q^2)&=O(1),
    &\ f_0^T(q^2)&=O(1),
    &\ f_g^T(q^2)&=O(1),\\
    f_\perp^{T5}(q^2)&=O(1),
    &\  f_0^{T5}(q^2)&=O(1),
    &\ f_g^{T5}(q^2)&=O(1),
\end{aligned}
\end{equation}
and the following constraints for $q^2\to (m_\Lb-m_\Lst)^2$
\begin{equation}\label{eq:endpointbehaviourT}
\begin{aligned}
    f_\perp^T(q^2)&=O\left(\textstyle\frac{1}{s_-}\right),
    &\ f_0^T(q^2)&=O\left(\textstyle\frac{1}{s_-}\right),
    &\ f_g^T(q^2)&=O(1),\\
    f_\perp^{T5}(q^2)&=O\left(\textstyle\frac{1}{\sqrt s_-}\right),
    &\  f_0^{T5}(q^2)&=O\left(\textstyle\frac{1}{\sqrt s_-}\right),
    &\ f_g^{T5}(q^2)&=O\left(\textstyle\frac{1}{\sqrt s_-}\right).
\end{aligned}
\end{equation}
These conditions are sufficient to ensure the absence of unphysical poles in the hadronic matrix elements, but once again, form factors exhibiting less singular behaviours are also acceptable.
Moreover, the equality $\sigma_{\mu\nu}\gamma_5=i\epsilon_{\mu\nu\rho\sigma}\sigma^{\rho\sigma}/2$
yields the following constraints for the values of the tensor form factors at $q^2=0$
\begin{equation}
    f_\perp^{T5}(0)=f_\perp^{T}(0),
         \qquad
  f_g^{T5}(0)=f_g^{T}(0)    \frac{m_\Lb+m_\Lst}{m_\Lb-m_\Lst}.
\end{equation}

As can be seen by comparing with the previous section, the situation is slightly different from the vector/axial case:
 there is no form factor corresponding to the time-like polarisation (or $q^\mu$), the normalisation of the Lorentz structures is different, and the resulting constraints at $q^2=0$ are different.

This leads to the helicity amplitudes
\begin{equation}
\begin{aligned}
H^T_t(s_{\Lb},s_\Lst)&\equiv\varepsilon^*_\mu(t)\braket{\Lst(k,s_\Lst) | \bar si\sigma^{\mu\nu}q_\nu b|\Lb(p,s_{\Lb})}=0,\\
H^T_0(s_{\Lb},s_\Lst)&\equiv\varepsilon^*_\mu(0)\braket{\Lst(k,s_\Lst) | \bar si\sigma^{\mu\nu}q_\nu b|\Lb(p,s_{\Lb})}\\
&=f_0^T(q^2)\frac{q^2}{s_+}\sqrt{\abs{e^2}}\bar u_\alpha(k,s_\Lst)p^\alpha u(p,s_{\Lb}),\\
H^T_\pm(s_{\Lb},s_\Lst)&\equiv\varepsilon^*_\mu(\pm)\braket{\Lst(k,s_\Lst) | \bar si\sigma^{\mu\nu}q_\nu b|\Lb(p,s_{\Lb})}\\
&=-\left(f_\perp^T(q^2) (m_{\Lb}+m_\Lst)+f_g^T(q^2)\frac{m_\Lst}{s_-}\right) \bar u_\alpha(k,s_\Lst)p^\alpha \slashed{\varepsilon}^*(\pm) u(p,s_{\Lb})\\
&\qquad- f_g^T(q^2) \bar u_\alpha(k,s_\Lst)\varepsilon^{*\alpha}(\pm) u(p,s_{\Lb}),\\
H^{T5}_t(s_{\Lb},s_\Lst)&\equiv\varepsilon^*_\mu(t)\braket{\Lst(k,s_\Lst) | \bar si\sigma^{\mu\nu}q_\nu \gamma^5 b|\Lb(p,s_{\Lb})}=0,\\
H^{T5}_0(s_{\Lb},s_\Lst)&\equiv\varepsilon^*_\mu(0)\braket{\Lst(k,s_\Lst) | \bar si\sigma^{\mu\nu}q_\nu \gamma^5 b|\Lb(p,s_{\Lb})}\\
&=f_0^{T5}(q^2)\frac{q^2}{s_-}\sqrt{\abs{e^2}}\bar u_\alpha(k,s_\Lst)\gamma^5 p^\alpha u(p,s_{\Lb}),\\
H^{T5}_\pm(s_{\Lb},s_\Lst)&\equiv\varepsilon^*_\mu(\pm)\braket{\Lst(k,s_\Lst) | \bar si\sigma^{\mu\nu}q_\nu \gamma^5 b|\Lb(p,s_{\Lb})}\\
&=\left(f_\perp^{T5}(q^2) (m_\Lb-m_\Lst)-f_g^{T5}(q^2)\frac{m_\Lst}{s_+}\right) \bar u_\alpha(k,s_\Lst)p^\alpha \slashed{\varepsilon}^*(\pm) \gamma^5 u(p,s_{\Lb})\\
&\qquad - f_g^{T5}(q^2) \bar u_\alpha(k,s_\Lst)\varepsilon^{*\alpha}(\pm) \gamma^5 u(p,s_{\Lb}).
\end{aligned}
\end{equation}
We recall that $e^\mu$ has been defined in Eq.~(\ref{eq:polarisations_general}).
As expected, there is no contribution from the time-like polarisation in the case of the tensor form factors. One obtains the following non-vanishing amplitudes
\begin{equation}
\begin{aligned}\label{eq:HplusHminus3}
H^T_0(+1/2,+1/2)=&H^T_0(-1/2,-1/2)=f_0^T(q^2)\sqrt{q^2}\frac{s_-\sqrt{s_+}}{\sqrt{6}m_\Lst},\\
H^T_+(+1/2,-1/2)=&H^T_-(-1/2,+1/2)=f_{\perp}^T(q^2)(m_{\Lb}+m_\Lst)\frac{s_-\sqrt{s_+}}{\sqrt{3}m_\Lst},\\
H^T_+(-1/2,-3/2)=&H^T_-(+1/2,+3/2)=-f_{g}^T(q^2)\sqrt{s_+},\\
H^{T5}_0(+1/2,+1/2)=&-H^{T5}_0(-1/2,-1/2)=-f_0^{T5}(q^2)\sqrt{q^2}\frac{s_+\sqrt{s_-}}{\sqrt{6}m_\Lst},\\
H^{T5}_+(+1/2,-1/2)=&-H^{T5}_-(-1/2,+1/2)=f_{\perp}^{T5}(q^2)(m_{\Lb}-m_\Lst)\frac{s_+\sqrt{s_-}}{\sqrt{3}m_\Lst},\\
H^{T5}_+(-1/2,-3/2)=&-H^{T5}_-(+1/2,+3/2)=-f_{g}^{T5}(q^2)\sqrt{s_-}.
\end{aligned}
\end{equation}

\subsection{$\Lb\to \Lst\ell^+\ell^-$ decay amplitudes}

Considering the effective Hamiltonian Eq.~(\ref{eq:hameff}) with only contributions
from $\C7,\C{9\ell},\C{10\ell}$ and their chirality-flipped counterparts and neglecting the lepton mass, we obtain the following decomposition:
\begin{eqnarray}
&&\mathcal{M}(s_{\Lb},s_{\Lst})\equiv N_1\braket{\Lst(s_{\Lst})\ell^+\ell^-|\sum_i\C{i}\mathcal{O}_i|\Lb(s_{\Lb})}\\
&&\quad =\frac{N_1}{2}\left\{\sum_{L(R)}L_{L(R)}^\mu\left[H^V_\mu \C{9,10,+}^{L(R)}-H^A_\mu \C{9,10,-}^{L(R)}-\frac{2m_b}{q^2}\left\{H^T_\mu(\C7+\C{7'})+H^{T5}_\mu(\C7-\C{7'})\right\}\right]\right\}, \nonumber
\end{eqnarray}
where the leptonic and hadronic helicity amplitudes read
\begin{equation}
L_{L(R)}^\mu=\bar{u}(k_2,s_{\ell^{-}})\gamma^\mu(1\pm \gamma_5) v(k_1,s_{\ell^+}),\qquad
H_\mu^X=\langle \Lst | \bar{s}\Gamma_\mu^X b | \Lb\rangle,
\end{equation}
with the various Dirac matrices $\Gamma_\mu^{V(A)}=\gamma_\mu(\gamma_5)$ and
$\Gamma_\mu^{T(T5)}=\sigma_{\mu\nu}q^\nu(\gamma_5)$.
The combinations of Wilson coefficients are defined as
\begin{equation}
\C{9,10,+}^{L(R)}=(\C{9\ell}\mp \C{10\ell})+(\C{9'\ell}\mp \C{10'\ell}), \qquad
\C{9,10,-}^{L(R)}=(\C{9\ell}\mp \C{10\ell})-(\C{9'\ell}\mp \C{10'\ell}),
\end{equation}
and the normalization reads
\begin{equation}
N_1=\frac{4G_F}{\sqrt{2}} V_{tb}V_{ts}^*\frac{\alpha}{4\pi}.
\end{equation}
We can perform the helicity amplitude decomposition discussed in Sec.~\ref{sec:effhamkin}, exploiting the expression of the hadronic helicity amplitudes in terms of the form factors described in Sec.~\ref{sec:hadronic} and using the explicit solutions of the Dirac equation in App.~\ref{app:kinematics} in order to determine the leptonic helicity amplitudes. The resulting expressions are given in Tab.~\ref{tab:LbtoLstAmpl}, with the corresponding hadronic transversity amplitudes
\begin{equation}\label{eq:ABampl}
\begin{aligned}
B_{\perp1}^{L(R)}=&+\sqrt{2}N\left(\C{9,10,+}^{L(R)}H_+^V(-1/2,-3/2)-\frac{2m_b(\C7+\C{7'})}{q^2}H_+^T(-1/2,-3/2)\right),\\
B_{\parallel1}^{L(R)}=&-\sqrt{2}N\left(\C{9,10,-}^{L(R)}H_+^A(-1/2,-3/2)+\frac{2m_b(\C7-\C{7'})}{q^2}H_+^{T5}(-1/2,-3/2)\right),\\
A_{\perp1}^{L(R)}=&+\sqrt{2}N\left(\C{9,10,+}^{L(R)}H_+^V(+1/2,-1/2)-\frac{2m_b(\C7+\C{7'})}{q^2}H_+^T(+1/2,-1/2)\right),\\
A_{\parallel1}^{L(R)}=&-\sqrt{2}N\left(\C{9,10,-}^{L(R)}H_+^A(+1/2,-1/2)+\frac{2m_b(\C7-\C{7'})}{q^2}H_+^{T5}(+1/2,-1/2)\right),\\
A_{\perp0}^{L(R)}=&+\sqrt{2}N\left(\C{9,10,+}^{L(R)}H_0^V(+1/2,+1/2)-\frac{2m_b(\C7+\C{7'})}{q^2}H_0^T(+1/2,+1/2)\right),\\
A_{\parallel0}^{L(R)}=&-\sqrt{2}N\left(\C{9,10,-}^{L(R)}H_0^A(+1/2,+1/2)+\frac{2m_b(\C7-\C{7'})}{q^2}H_0^{T5}(+1/2,+1/2)\right).
\end{aligned}
\end{equation}
The normalisation factor $N$, related to the 4-body phase space of this decay, is defined as
\begin{equation}\label{eq:NormalisationN}
N=N_1\sqrt{\frac{q^2\sqrt{\lambda(m_\Lb^2,m_\Lst^2,q^2)}}{3\cdot 2^{10} m_\Lb^3\pi^3}}.
\end{equation}
We have used the relations Eqs.~(\ref{eq:HplusHminus1}), (\ref{eq:HplusHminus2}), (\ref{eq:HplusHminus3}) in order to express $H_-$ amplitudes in terms of $H_+$. We notice that there are no contributions from $H_t$ here: the tensor hadronic amplitudes vanish exactly, whereas the vector/axial hadronic amplitudes are multiplied by the leptonic helicity amplitude $\epsilon^\mu (t)L_{\mu,L(R)}$ which are proportional to $m_\ell^2/\sqrt{q^2}$ (and neglected here) due to the lepton equation of motion.

\begin{table}
{\renewcommand{\arraystretch}{2}%
$$\begin{array}{c|c|c}
 s_{\Lb}&s_{\Lst}&\mathcal{M}\\ \hline
+\frac{1}{2}&+\frac{1}{2}& \displaystyle \frac{N_1}{2\sqrt{2}N}\sum_{L(R)}\left[A_{\perp0}^{L(R)}+A_{\parallel0}^{L(R)} \right]\bar u \slashed {\varepsilon}(0)P_{L(R)}v \\ \hline
-\frac{1}{2}&-\frac{1}{2}& \displaystyle \frac{N_1}{2\sqrt{2}N}\sum_{L(R)}\left[A_{\perp0}^{L(R)}-A_{\parallel0}^{L(R)} \right]\bar u \slashed {\varepsilon}(0)P_{L(R)}v \\ \hline
+\frac{1}{2}&-\frac{1}{2}& \displaystyle \frac{N_1}{2\sqrt{2}N}\sum_{L(R)}\left[A_{\perp1}^{L(R)}+A_{\parallel1}^{L(R)} \right]\bar u \slashed {\varepsilon}(+)P_{L(R)}v \\ \hline
-\frac{1}{2}&+\frac{1}{2}& \displaystyle \frac{N_1}{2\sqrt{2}N}\sum_{L(R)}\left[A_{\perp1}^{L(R)}-A_{\parallel1}^{L(R)} \right]\bar u \slashed {\varepsilon}(-)P_{L(R)}v\\ \hline
-\frac{1}{2}&-\frac{3}{2}& \displaystyle \frac{N_1}{2\sqrt{2}N}\sum_{L(R)}\left[B_{\perp1}^{L(R)}+B_{\parallel1}^{L(R)} \right]\bar u \slashed {\varepsilon}(+)P_{L(R)}v \\ \hline
+\frac{1}{2}&+\frac{3}{2}& \displaystyle \frac{N_1}{2\sqrt{2}N}\sum_{L(R)}\left[B_{\perp1}^{L(R)}-B_{\parallel1}^{L(R)} \right]\bar u \slashed {\varepsilon}(-)P_{L(R)}v
\end{array}$$}
\caption{$\Lb\to \Lst\ell^+\ell^-$ decay amplitudes in terms of hadronic transversity amplitudes}
\label{tab:LbtoLstAmpl}
\end{table}

\subsection{$\Lst\to N\bar{K}$ decay}

The $\Lst\to N\bar{K}$ decay rate~\footnote{We have not been specific whether we perform the sum over the two isospin states or select only one of them. This has no impact on the computation as long as the same definition is used for $\Lst\to N\bar{K}$ and $\Lb\to \Lst(\to N\bar{K})\ell^+\ell^-$.} can be computed using
\begin{equation}
\Gamma(\Lst\to N\bar{K}) = \frac{\beta_{N\bar{K}}}{16\pi m_{\Lst}}{|\overline{\mathcal{M}^\Lst}|}^2,
\end{equation}
We can consider either of the two interaction terms discussed in Sec.~\ref{sec:lambdastar}, corresponding to
\begin{equation}
\begin{aligned}
\mathcal{M}^\Lst_1(m,s)=& gm_\Lst  k_2^{\mu} \bar{u}^s \gamma_5 U_\mu^m,\\
\mathcal{M}^\Lst_2(m,s)=& g\varepsilon^{\mu\nu\alpha\beta}k_\mu k_{2\beta}\bar{u}^s \gamma_\alpha U_\nu^m,
\end{aligned}
\end{equation}
These two alternative choices for the interaction terms describe the same physical decay for on-shell particles, and we checked explicitly that these two choices are equivalent and lead to the same final results in the following.
From Eqs.~(\ref{eq:M}) and (\ref{eq:dGamma}), we see that the computation of the $\Lb\to \Lst(\to N\bar{K})\ell^+\ell^-$ decay rate will require the interference terms between matrix elements with different $\Lst$ polarisations, which can be defined as
\begin{equation}\label{eq:Gamma2}
\Gamma_2(s_\Lst^a,s_\Lst^b)\equiv \frac{\sqrt{r_+r_-}}{16\pi m_\Lst^3}\sum_{s_N}\mathcal{M}(s_\Lst^a,s_N)\mathcal{M}(s_\Lst^b,s_N)^*,
\end{equation}
where $r_\pm=(m_{\Lst}^2\pm m_N^2)-m_{\bar{K}}^2$.
The normalisation for $\Gamma_2$ in Eq.~(\ref{eq:Gamma2}) comes from the phase space, which is present both in $\Lst\to N\bar{K}$ and $\Lb\to \Lst(\to N\bar{K})\ell^+\ell^-$.
This definition is such that the $\Lst\to N\bar{K}$ decay reads
\begin{equation}
\Gamma(\Lst\to N\bar{K}) = \sum_{s_\Lst}\frac{\Gamma_2(s_\Lst,s_\Lst)}{4}.
\end{equation}
Using the explicit expression of the solutions in App.~\ref{app:kinematics}, we obtain
\begin{equation}\footnotesize
\Gamma_2=\frac{\mathcal{B}_\Lst \Gamma_\Lst}{4}\left(
\begin{array}{cccc}
 6 \sin ^2(\theta_\Lst ) & 2 \sqrt{3} e^{-i \phi } \sin (2 \theta_\Lst ) &
   -2 \sqrt{3} e^{-2 i \phi } \sin ^2(\theta_\Lst ) & 0 \\
 2 \sqrt{3} e^{i \phi } \sin (2 \theta_\Lst ) & 3 \cos (2 \theta_\Lst )+5 &
   0 & -2 \sqrt{3} e^{-2 i \phi } \sin ^2(\theta_\Lst ) \\
 -2 \sqrt{3} e^{2 i \phi } \sin ^2(\theta_\Lst ) & 0 & 3 \cos (2 \theta_\Lst
   )+5 & -2 \sqrt{3} e^{-i \phi } \sin (2 \theta_\Lst ) \\
 0 & -2 \sqrt{3} e^{2 i \phi } \sin ^2(\theta_\Lst ) & -2 \sqrt{3} e^{i \phi }
   \sin (2 \theta_\Lst ) & 6 \sin ^2(\theta_\Lst ) \\
\end{array}
\right),
\end{equation}
with rows and columns corresponding to values of $s_a,s_b=-3/2,-1/2,1/2,3/2$. We denote $\mathcal{B}_\Lst \equiv\mathcal{B}(\Lst\to K^-p)=\mathcal{B}(\Lst\to \bar{K}^0n)$ and $\Gamma_\Lst$ is the inclusive decay width of the $\Lst$ baryon.

\section{Phenomenology}\label{sec:pheno}

\subsection{Angular observables}\label{sec:angular}

Combining all the above elements, we obtain finally the differential decay rate
\begin{equation}\label{eq:angobs}
\begin{aligned}
L(q^2,\theta_\ell,\theta_\Lst,\phi)=&\frac{8\pi}{3}\frac{d^4\Gamma}{dq^2d\cos{\theta_\ell}d\cos{\theta_\Lst}d\phi}\\
=&\cos ^2\theta_\Lst \left(L_{1c} \cos \theta_\ell+L_{1cc} \cos ^2\theta_\ell+L_{1ss} \sin ^2\theta_\ell\right)\\
&+ \sin ^2\theta_\Lst \left(L_{2c} \cos
   \theta_\ell+L_{2cc} \cos ^2\theta_\ell+L_{2ss} \sin ^2\theta_\ell\right)\\
&+ \sin ^2\theta_\Lst \left(L_{3ss} \sin ^2\theta_\ell \cos^2
   \phi+L_{4ss} \sin ^2\theta_\ell \sin \phi \cos
   \phi\right)\\
   &+\sin \theta_\Lst \cos \theta_\Lst \cos \phi (L_{5s} \sin \theta_\ell+L_{5sc} \sin \theta_\ell \cos \theta_\ell)\\
   &+\sin \theta_\Lst \cos \theta_\Lst\sin \phi (L_{6s} \sin
   \theta_\ell+L_{6sc} \sin \theta_\ell \cos \theta_\ell),
\end{aligned}
\end{equation}
with the angular coefficients $L$ that are interferences between the various helicity amplitudes defined in Tab.~\ref{tab:LbtoLstAmpl}:
\begin{equation}
\begin{aligned}
L_{1c}=&-2\mathcal{B}_\Lst\left(\textrm{Re}(A_{\perp1}^{L}A_{\parallel1}^{L*})-(L\leftrightarrow R)\right),\\
 L_{1cc}=&\mathcal{B}_\Lst\left(\left|
   A_{\parallel1}^{L}\right| ^2+\left| A_{\perp1}^{L}\right| ^2+(L\leftrightarrow R)\right),\\
 L_{1ss}=&\frac{1}{2} \mathcal{B}_\Lst\left(2 \left|
   A_{\parallel0}^{L}\right| ^2+2 \left|
   A_{\perp0}^{L}\right| ^2+\left| A_{\parallel1}^{L}\right| ^2+\left| A_{\perp1}^{L}\right| ^2+(L\leftrightarrow R)\right),\\
 L_{2c}=&-\frac{1}{2} \mathcal{B}_\Lst
   \left(\textrm{Re}(A_{\perp1}^{L} A_{\parallel1}^{L*})+3 \textrm{Re}(B_{\perp1}^{L} B_{\parallel1}^{L*})
   -(L\leftrightarrow R)\right),\\
 L_{2cc}=&\frac{1}{4} \mathcal{B}_\Lst
   \left(\left| A_{\parallel1}^{L}\right| ^2+\left| A_{\perp1}^{L}\right|^2+3
   \left| B_{\parallel1}^{L}\right| ^2+3\left| B_{\perp1}^{L}\right|^2+(L\leftrightarrow R)\right),\\
   L_{2ss}=&\frac{1}{8} \mathcal{B}_\Lst \left[2
   \left| A_{\parallel0}^{L}\right| ^2+\left| A_{\parallel1}^{L}\right| ^2+2 \left|
   A_{\perp0}^{L}\right| ^2+\left| A_{\perp1}^{L}\right| ^2+3 \left|
   B_{\parallel1}^{L}\right| ^2+3 \left| B_{\perp1}^{L}\right| ^2\right.\\
   & \qquad\qquad \left.
   -2\sqrt{3}\textrm{Re}(B_{\parallel1}^{L} A_{\parallel1}^{L*})+2\sqrt{3}\textrm{Re}(B_{\perp1}^{L}
   A_{\perp1}^{L*})+(L\leftrightarrow R)\right],\\
 L_{3ss}=&\frac{\sqrt{3}}{2}
   \mathcal{B}_\Lst\left(\textrm{Re}(B_{\parallel1}^{L}
   A_{\parallel1}^{L*})-\textrm{Re}(B_{\perp1}^{L} A_{\perp1}^{L*})+(L\leftrightarrow R)\right),\\
L_{4ss}=&\frac{\sqrt{3}}{2}
   \mathcal{B}_\Lst\left(\textrm{Im}(B_{\perp1}^{L}
   A_{\parallel1}^{L*})-\textrm{Im}(B_{\parallel1}^{L} A_{\perp1}^{L*})
   + (L\leftrightarrow R)\right),\\
L_{5s}=&
   \sqrt{\frac{3}{2}} \mathcal{B}_\Lst\left(
   \textrm{Re}(B_{\perp1}^{L}
   A_{\parallel0}^{L*})- \textrm{Re}(B_{\parallel1}^{L} A_{\perp0}^{L*})
   -(L\leftrightarrow R)\right),\\
L_{5sc}=& \sqrt{\frac{3}{2}}
   \mathcal{B}_\Lst\left(-\textrm{Re}(B_{\parallel1}^{L}
   A_{\parallel0}^{L*})+\textrm{Re}(B_{\perp1}^{L} A_{\perp0}^{L*})
   +(L\leftrightarrow R)\right),\\
L_{6s}=& \sqrt{\frac{3}{2}}
   \mathcal{B}_\Lst\left(\textrm{Im}(B_{\parallel1}^{L}
   A_{\parallel0}^{L*})-\textrm{Im}(B_{\perp1}^{L} A_{\perp0}^{L*})
   -(L\leftrightarrow R)
    \right),\\
L_{6sc}=&- \sqrt{\frac{3}{2}}
   \mathcal{B}_\Lst\left(
   \textrm{Im}(B_{\perp1}^{L}A_{\parallel0}^{L*})-\textrm{Im}(B_{\parallel1}^{L} A_{\perp0}^{L*})
   +(L\leftrightarrow R)\right),
\end{aligned}
\end{equation}
where we have neglected the lepton masses. The corresponding CP-conjugate mode will involve $\bar{A}$ and $\bar{B}$ amplitudes, where only the weak phases are taken to their opposite, as already discussed in Sec.~\ref{sec:kinematics}.

We provide further cross-checks of these expressions in App.~\ref{app:crosscheckang} by comparing our results with general expectations from the partial-wave analysis of four-body $b\to s\ell^+\ell^-$ decays~\cite{Gratrex:2015hna}, and
in App.~\ref{app:crosscheckgamma} by checking the agreement with the expressions for $\Lb\to\Lst(\to KN)\gamma$~\cite{Hiller:2007ur,Legger:2006cq}.

\subsection{Derived observables}

One can define derived observables using a particular weight $\omega$ to integrate the differential decay rate over the whole phase space
\begin{equation}
X[\omega](q^2)\equiv\int\frac{d^4\Gamma}{dq^2d\cos{\theta_\ell}d\cos{\theta_\Lst}d\phi}\omega(q^2,\theta_\ell,\theta_\Lst,\phi)d\cos{\theta_\ell}d\cos{\theta_\Lst}d\phi.
\end{equation}

The differential decay width is
\begin{equation}
\begin{aligned}
\frac{d\Gamma}{dq^2}&=X[1]=\frac{1}{3}[L_{1cc}+2L_{1ss}+2L_{2cc}+4L_{2ss}+2L_{3ss}]\\
   &=|A_{||0}^L|^2+|A_{\perp 0}^L|^2+|A_{||1}^L|^2+|A_{\perp 1}^L|^2+|B_{||1}^L|^2+|B_{\perp 1}^L|^2+(L\leftrightarrow R),
   \end{aligned}
\end{equation}
which we can use to normalise the CP-averaged angular observables and the corresponding CP-asymmetries
\begin{equation}\label{eq:SandA}
S_{i}=\frac{L_i+\bar{L}_i}{d(\Gamma+\bar\Gamma)/dq^2}, \qquad A_{i}=\frac{L_i-\bar{L}_i}{d(\Gamma+\bar\Gamma)/dq^2}.
\end{equation}

One can similarly define the transverse and longitudinal polarization of the dilepton system~\cite{Boer:2014kda}
 \begin{equation}
 \begin{aligned}
 F_1=&X\left[\frac{5\cos\theta^2_\ell-1}{d\Gamma/dq^2}\right]=\frac{2 (L_{1cc}+2 L_{2cc})}{L_{1cc}+2 (L_{1ss}+L_{2cc}+2 L_{2ss}+L_{3ss})},\\
  F_0=&1-F_1=1-\frac{2 (L_{1cc}+2 L_{2cc})}{L_{1cc}+2 (L_{1ss}+L_{2cc}+2 L_{2ss}+L_{3ss})}.
\end{aligned}
  \end{equation}
One can also define a
 forward-backward asymmetry with respect to the leptonic scattering angle normalised to the differential rate
 \begin{equation}
 A^\ell_{FB}=X\left[\frac{\text{sgn}[\cos\theta_\ell]}{d\Gamma/dq^2}\right]=\frac{3 (L_{1c}+2 L_{2c})}{2 (L_{1cc}+2 (L_{1ss}+L_{2cc}+2 L_{2ss}+L_{3ss}))}.
 \end{equation}

Due to the strong decay of the $\Lst$, it is no surprise that the analogous asymmetries for the hadronic system vanish
 \begin{equation}
 A^\Lst_{FB}=X\left[\frac{\text{sgn}[\cos\theta_\Lst]}{d\Gamma/dq^2}\right]=0,\qquad
 A^{\ell\Lst}_{FB}=X\left[\frac{\text{sgn}[\cos\theta_\Lst\cos\theta_\ell]}{d\Gamma/dq^2}\right]=0.
 \end{equation}
These relations can be used in the context of an experimental analysis as simple tests of the correct identification of the $\Lst$ baryon within the $\Lambda\to N\bar{K}\ell^+\ell^-$ sample.

\subsection{Low- and large-recoil limits}

As can be seen from the previous expressions, the description of this decay involves 8 vector/axial form factors and 6 tensor form factors. This number is considerably reduced in the heavy quark limit $m_b\to\infty$. Two different kinematic situations can be considered: either the outgoing $\Lst$ baryon is soft (low-recoil limit) or it is energetic (large-recoil limit). Two different effective theories have been devised to exploit the hierarchy of soft and hard scales in both configurations, namely the Heavy Quark Effective Theory (HQET)~\cite{Isgur:1989vq,Isgur:1989ed,Isgur:1990pm,Mannel:1990vg,Grinstein:2004vb} and the Soft-Collinear Effective Theory (SCET)~\cite{Charles:1998dr,Beneke:2000wa,Bauer:2000yr,Beneke:2001at}.

In the low-recoil limit where HQET is valid~\cite{Mannel:1990vg}, one can use the heavy-baryon velocity $v^\mu=p^\mu/m_\Lb$ to project the
the $b$-quark field onto its large-spinor component $h_v=\slashed v h_v$:
\begin{equation}\label{eq:HQET}
\langle \Lst| \bar{s}\Gamma b|\Lb\rangle \to \bar{u}_\Lst^\alpha v_\alpha [\zeta_1+\slashed{v}\zeta_2]\Gamma u_\Lb,
\end{equation}
where $\Gamma$ is any Dirac matrix, $\zeta_1$ and $\zeta_2$ are the only two form factors that should be present  at leading order in $\alpha_S$ and $\Lambda/m_b$ according to HQET. These two form factors are functions of $q^2$ or equivalently $v\cdot v'$ (where $v'=k/m_\Lst$ is the velocity of the light baryon).
We can take the heavy-quark limit (neglecting $\Lambda/m_b$ contributions) in the definition of the form factors Eqs.~(\ref{eq:ffVA}) and (\ref{eq:ffT}) and identify the results with Eq.~(\ref{eq:HQET}). This is performed (with slightly different definitions of the form factors)
in Refs.~\cite{Mannel:2011xg,Feldmann:2011xf,Boer:2014kda}, and the corresponding expressions yield at low recoil:
\begin{equation}\label{eq:ffHQET}
\begin{aligned}
 f_\perp^V= f_0^V= f_t^A=f_\perp^T=f_0^T=&(\zeta_1-\zeta_2)/m_\Lb,\\
 f_\perp^A= f_0^A= f_t^V=f_\perp^{T5}=f_0^{T5}=&(\zeta_1+\zeta_2)/m_\Lb,\\
f_g^V=f_g^A=f_g^T=f_g^{T5}=&0.
\end{aligned}
\end{equation}
It is also possible to perform a similar analysis in the large-recoil limit where SCET holds. Following Ref.~\cite{Mannel:2011xg,Feldmann:2011xf,Boer:2014kda}, one can see that the SCET analysis yields:
\begin{equation} \label{eq:SCET}
\langle \Lst | \bar{s}\Gamma b|\Lb\rangle \to \xi \bar{u}_\Lst^\alpha v_\alpha \Gamma u_\Lb,
\end{equation}
where $\Gamma$ is any Dirac matrix, $\xi$ is the only form factor that should be present  at leading order in $\alpha_S$ and $\Lambda/m_b$ according to SCET. These form factors are functions of $q^2$ or equivalently $n_+\cdot k$ (where $n_+$ is a light-like vector orthogonal to $k$). One can see that formally, the expression for SCET Eq.~(\ref{eq:SCET}) will be obtained from the HQET expression Eq.~(\ref{eq:HQET}) by identifying $\zeta_1$ to $\xi$ and setting $\zeta_2$ to 0, leading to the equality at large recoil:
\begin{equation}\label{eq:ffSCET}
 f_t^V= f_\perp^V= f_0^V= f_t^A= f_\perp^A= f_0^A=f_\perp^T=f_0^T=f_\perp^{T5}=f_0^{T5}=\xi/m_\Lb,
\end{equation}
whereas all $f_g$ form factors vanish also in the large-recoil limit.

The methods used in Refs.~\cite{Mannel:2011xg,Feldmann:2011xf} could be used to analyse higher-order corrections to these relations (in powers of $\alpha_S$ and $\Lambda/m_b$) but this is out of the scope of the present article. From Eq.~(\ref{eq:ABampl}), we see that all the hadronic amplitudes $A_\perp$ involve $\zeta_1-\zeta_2$, whereas  $A_{||}$ involve $\zeta_1+\zeta_2$. All $B_\perp$ and $B_{||}$ amplitudes vanish in both limits because they only depend on $f_g$ form factors. Using the equalities in Eq.~(\ref{eq:ffHQET}) (which are also compatible with the equalities in Eq.~(\ref{eq:ffSCET})), we obtain
\begin{equation}\label{eq:K-HQET}
\begin{aligned}
 L_{1c}&\to \alpha (\zeta_1^2-\zeta_2^2),\qquad& L_{2c}&\to \frac{1}{4}L_{1c},  \\
 L_{1cc}&\to \alpha' (\zeta_1-\zeta_2)^2+\beta' (\zeta_1+\zeta_2)^2, \qquad&L_{2cc}&\to \frac{1}{4}L_{1cc},  \\
 L_{1ss}&\to \alpha'' (\zeta_1-\zeta_2)^2+\beta'' (\zeta_1+\zeta_2)^2,\qquad&
 L_{2ss}&\to \frac{1}{4}L_{1ss},
 \end{aligned}
\end{equation}
whereas the rest of the angular observables ($ L_{3ss},L_{4ss},L_{5s},L_{5sc},L_{6s},L_{6sc}$) vanishes.
The $\alpha$ and $\beta$ coefficients combine Wilson coefficients and kinematic factors. Considering the relations in Eqs.~(\ref{eq:K-HQET}), we can see that we cannot build in a straightforward manner optimised observables similar to the $B\to K^*\ell^+\ell^-$ channel~\cite{Matias:2012xw,Descotes-Genon:2013vna} where the form factors will cancel out and non-trivial information on the Wilson coefficients can be obtained (up to $1/m_b$ and $\alpha_S$ corrections).

In the case of the large-recoil limit, the three independent observables $ L_{1c}, L_{1cc}, L_{1ss}$ only depend on $\xi$ and any ratio of these observables for which the uncertainties coming from the form factors are suppressed by $1/m_b$.

This discussion leads us not to consider further the possibility of optimised observables and to focus on the normalised CP-averaged angular observables $S$.

\subsection{Numerical illustrations}\label{sec:numerical}

We consider now numerical results for the various angular coefficients described above. This should be considered as a preliminary study, as we are going to make several simplifications that should be reassessed carefully if one wants to provide accurate predictions for this decay.

Indeed, a complete analysis would require a precise knowledge of the 14 form factors described in Sec.~\ref{sec:hadronic} and their correlations. Preliminary lattice determinations have been presented in Ref.~\cite{Meinel:2016cxo} and the question of non-local contributions could be tackled data-driven methods similar to Refs.~\cite{Bobeth:2017vxj,Blake:2017fyh}, involving light-cone sum rules similar to Refs.~\cite{Feldmann:2011xf,Khodjamirian:2010vf,Khodjamirian:2012rm,Wang:2015ndk}. Since these results are not available yet, we will present a numerical illustration based on the MCN quark model of Ref.~\cite{Mott:2011cx}, in order to get an idea of the sensitivity of the angular coefficients to different NP scenarios. Let us add that the results of Ref.~\cite{Mott:2011cx} obey rather well the HQET relations Eq.~(\ref{eq:ffHQET}), but do not follow the SCET expectations very well Eq.~(\ref{eq:ffSCET}).

We focus on the muon case ($\ell=\mu$ in the following) and on the contributions coming from $O_{7,9,10,7',9',10'}$. We also add contributions from the other operators $O_{1-6},O_{8g}$ (using the values tabulated in Ref.~\cite{Descotes-Genon:2013vna}), but for simplicity, we consider only the factorisable quark-loop contributions coming from these operators, which can be included into effective Wilson coefficients ${\cal C}_7^{\rm eff}$ and ${\cal C}_9^{\rm eff}(q^2)$~\cite{Beneke:2000wa,Beneke:2001at}. This means that for the latter, we consider only the charm-loop contributions derived from perturbation theory (taking $m_c=1.3$ GeV), as there are no estimates for the long-distance contributions, contrary to $B\to K(^*)\ell^+\ell^-$ where several estimates based on different theoretical approaches are available~\cite{Khodjamirian:2010vf,Khodjamirian:2012rm,Bobeth:2017vxj,Blake:2017fyh,Ciuchini:2017mik,Capdevila:2017ert,Arbey:2018ics}.

\begin{figure}
    \centering
    \includegraphics[width=12cm]{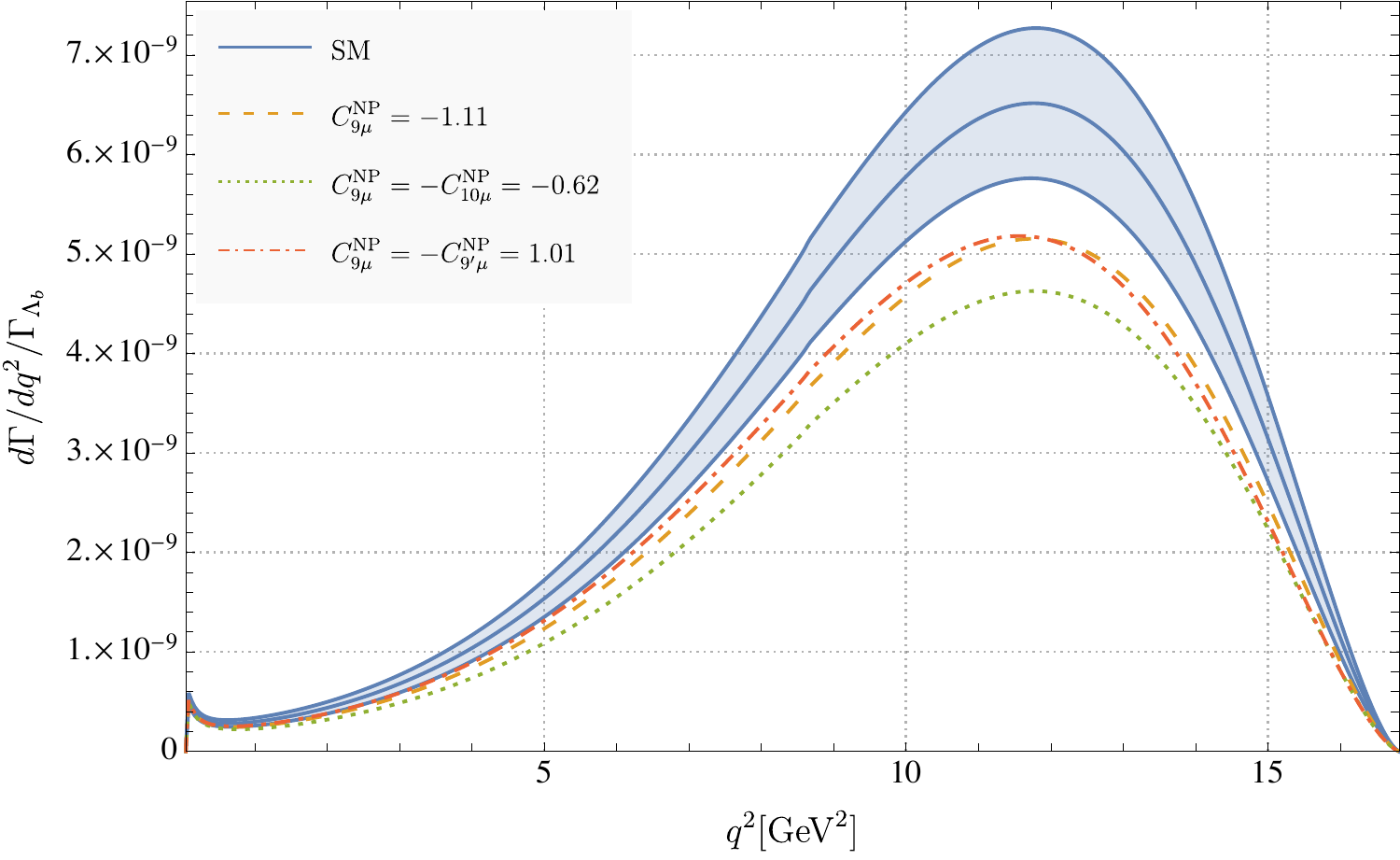}
    \includegraphics[width=12cm]{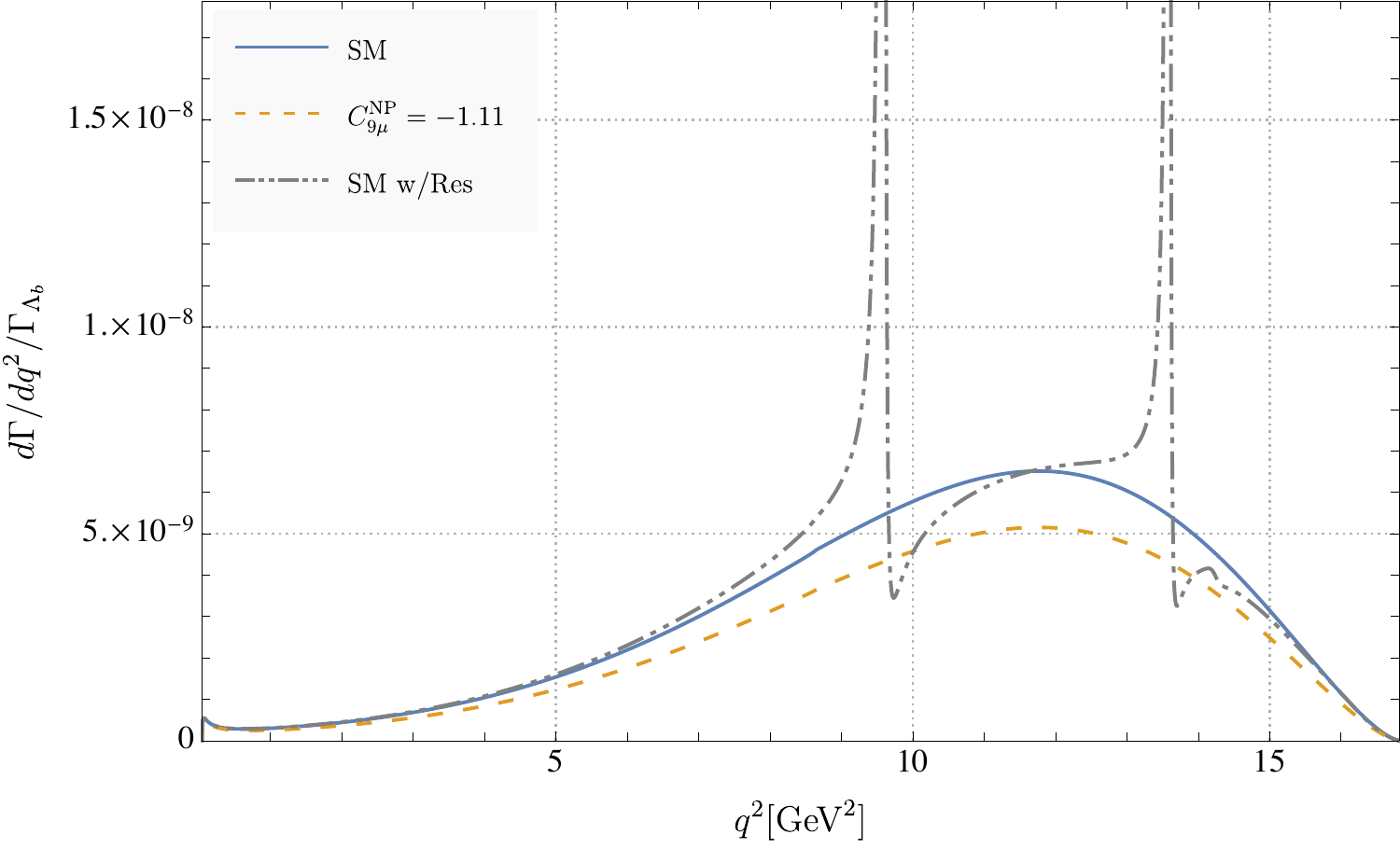}
    \caption{Top: Differential decay rate (normalised to the total $\Lb$ decay width) in the SM case and three NP scenarios. Only short-distance contributions from charm loops are include. Bottom: For illustration only, we also show the effect of a model for charmonium resonances in the SM case~\cite{Mott:2011cx}.}
    \label{fig:decayrate}
\end{figure}

\begin{figure}
    \centering
    \includegraphics[width=12cm]{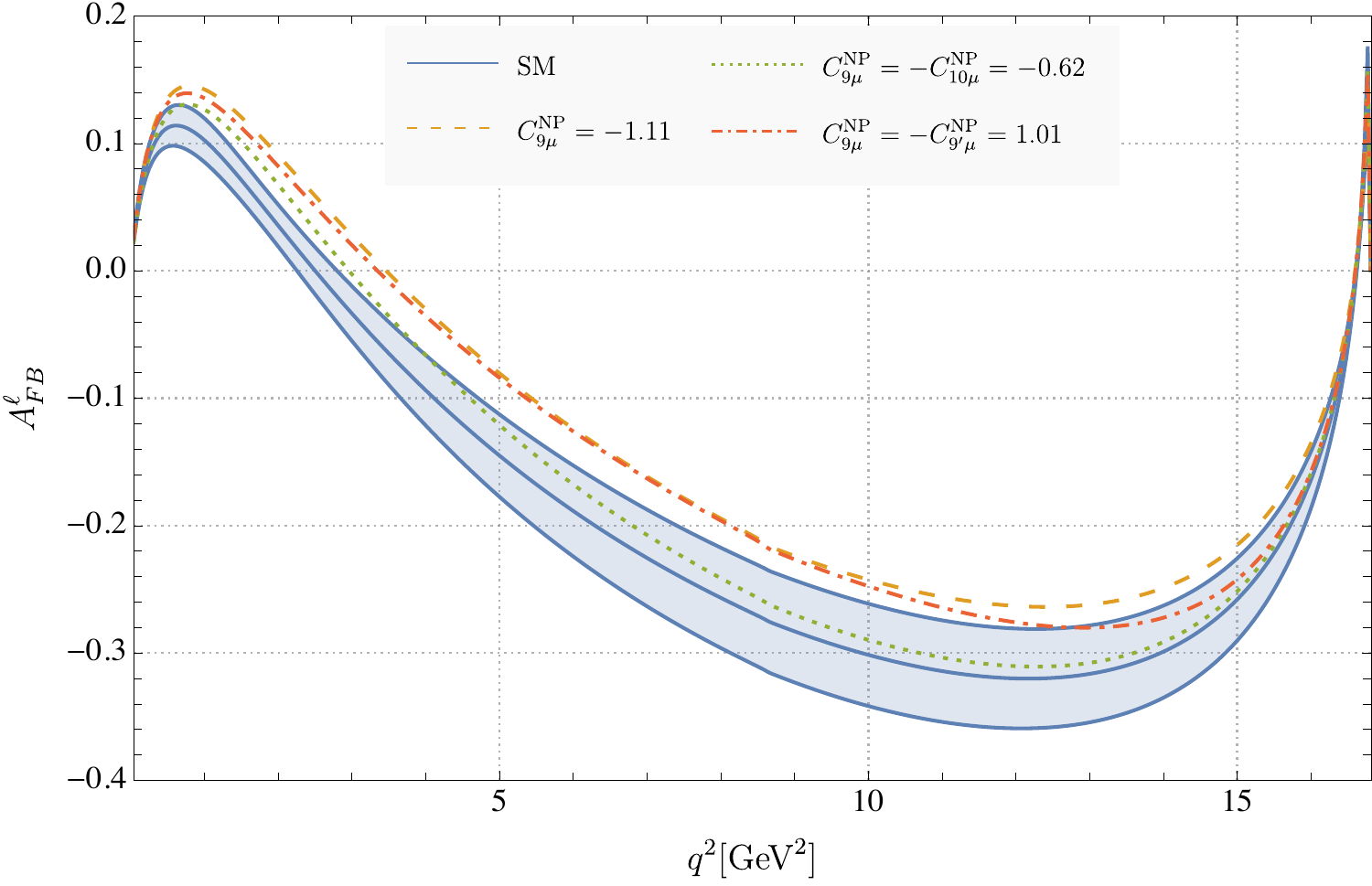}
    \includegraphics[width=12cm]{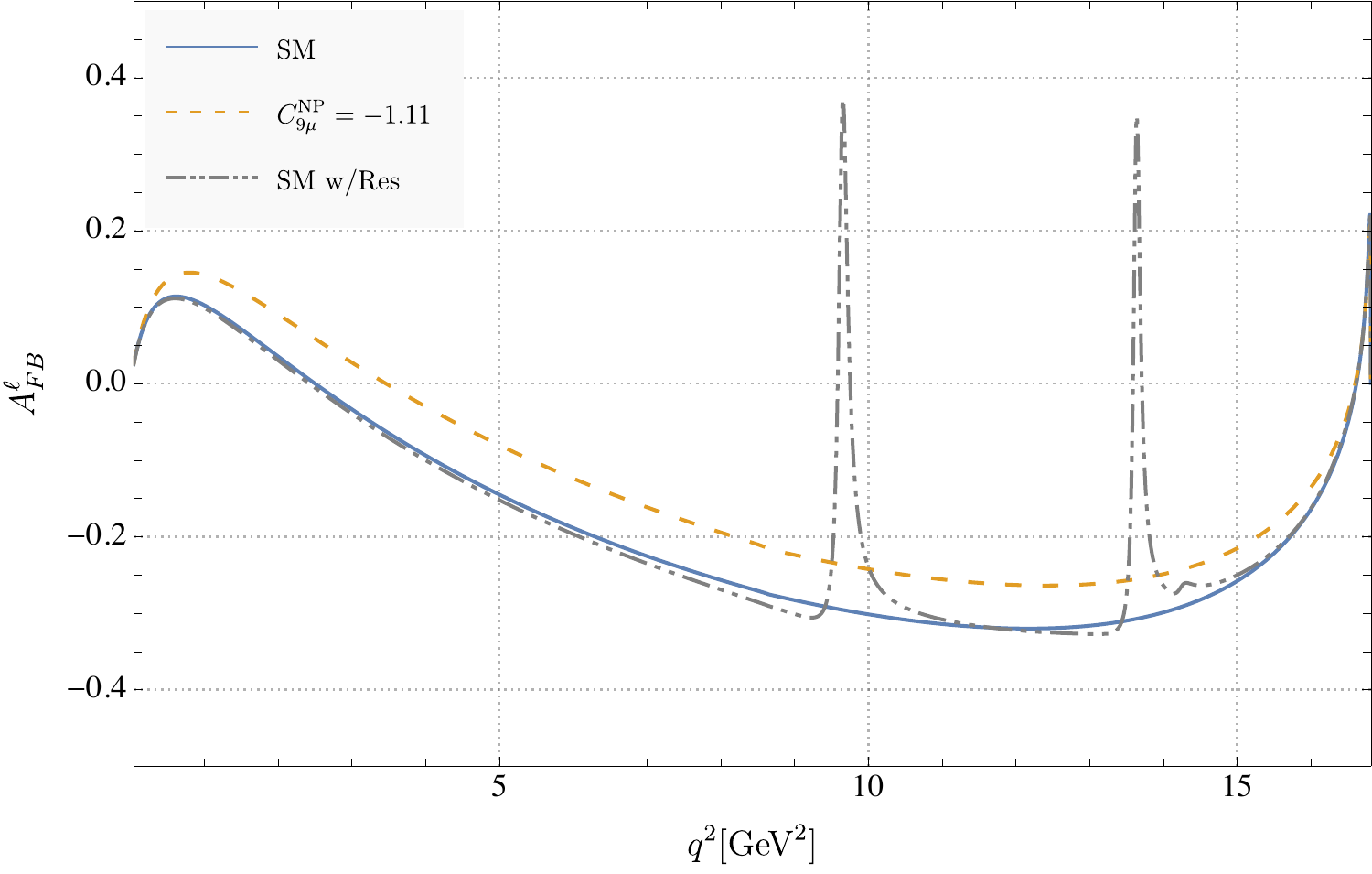}
    \caption{Top: $A^\ell_{FB}$ in the SM case and three NP scenarios. Only short-distance contributions from charm loops are include. Bottom: For illustration only, we also show the effect of a model for charmonium resonances in the SM case~\cite{Mott:2011cx}.}
    \label{fig:AFB}
\end{figure}

\begin{figure}
    \centering
    \includegraphics[width=12cm]{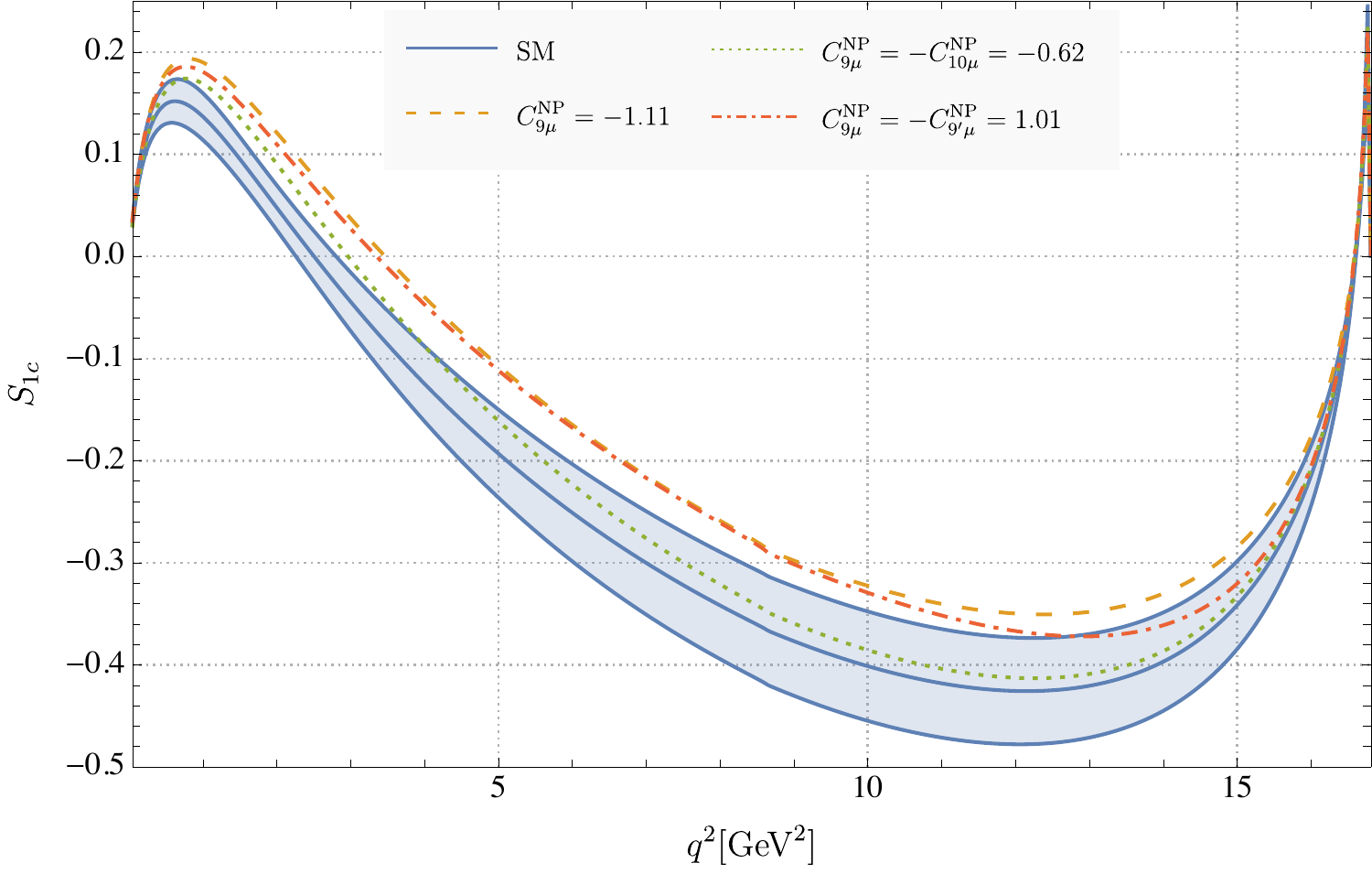}
    \includegraphics[width=12cm]{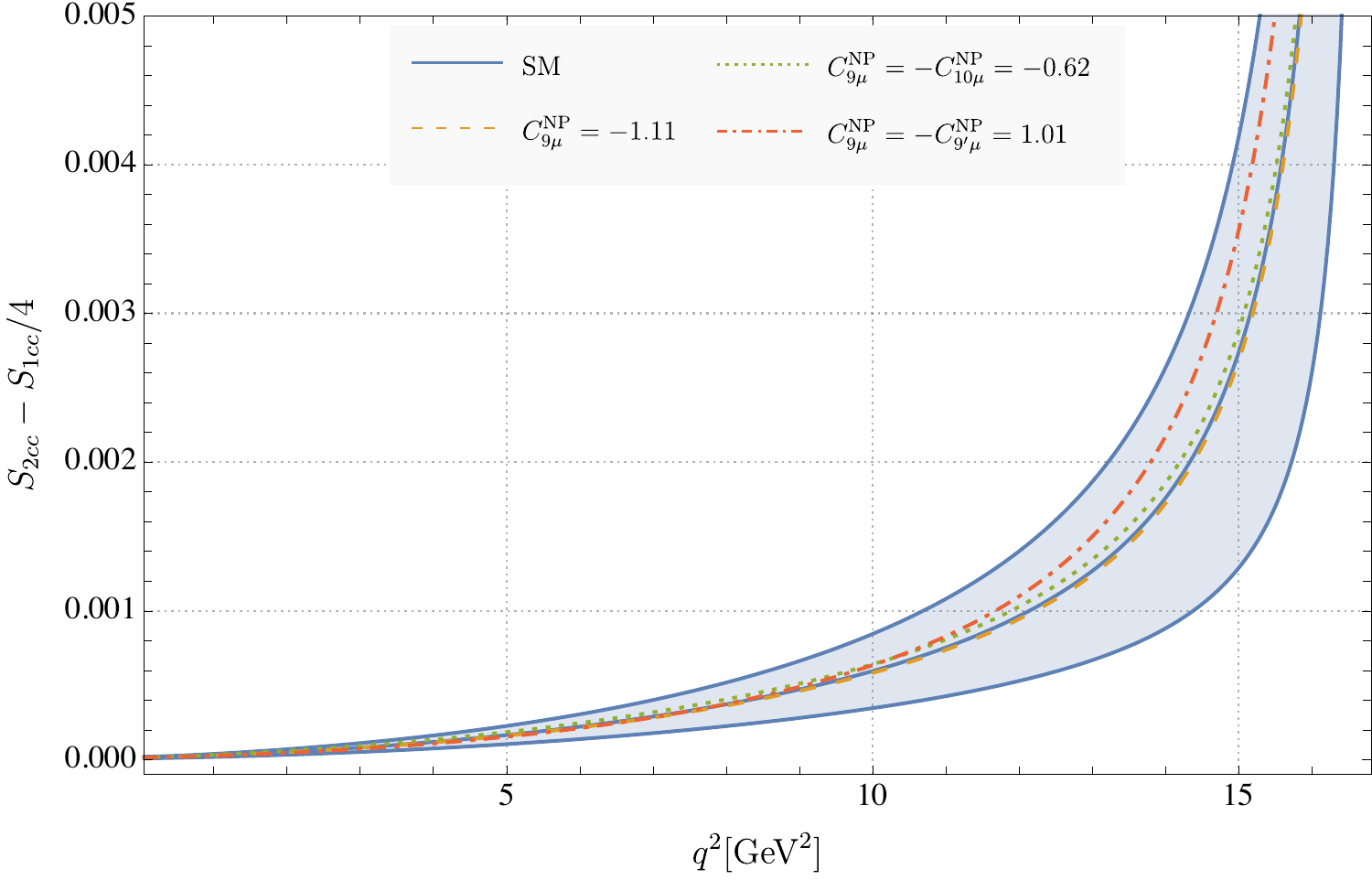}
    \caption{Variation of $S_{1c}$ (top) and $S_{2cc}-S_{1cc}/4$ (bottom) with respect to the dilepton invariant mass, in the case of the SM and three NP scenarios.}
    \label{fig:Sfig1}
\end{figure}

\begin{figure}
    \centering
    \includegraphics[width=12cm]{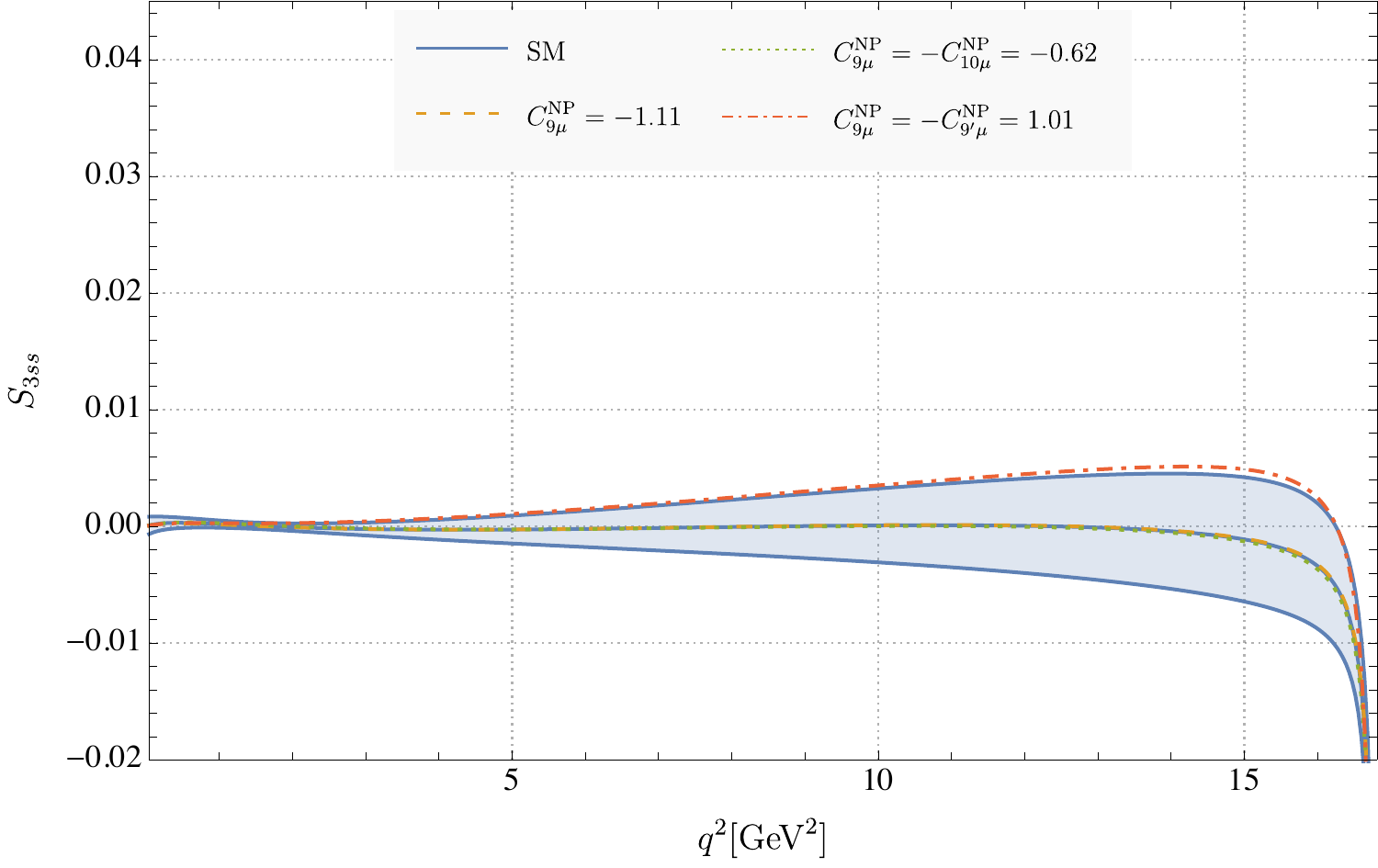}
    \includegraphics[width=12cm]{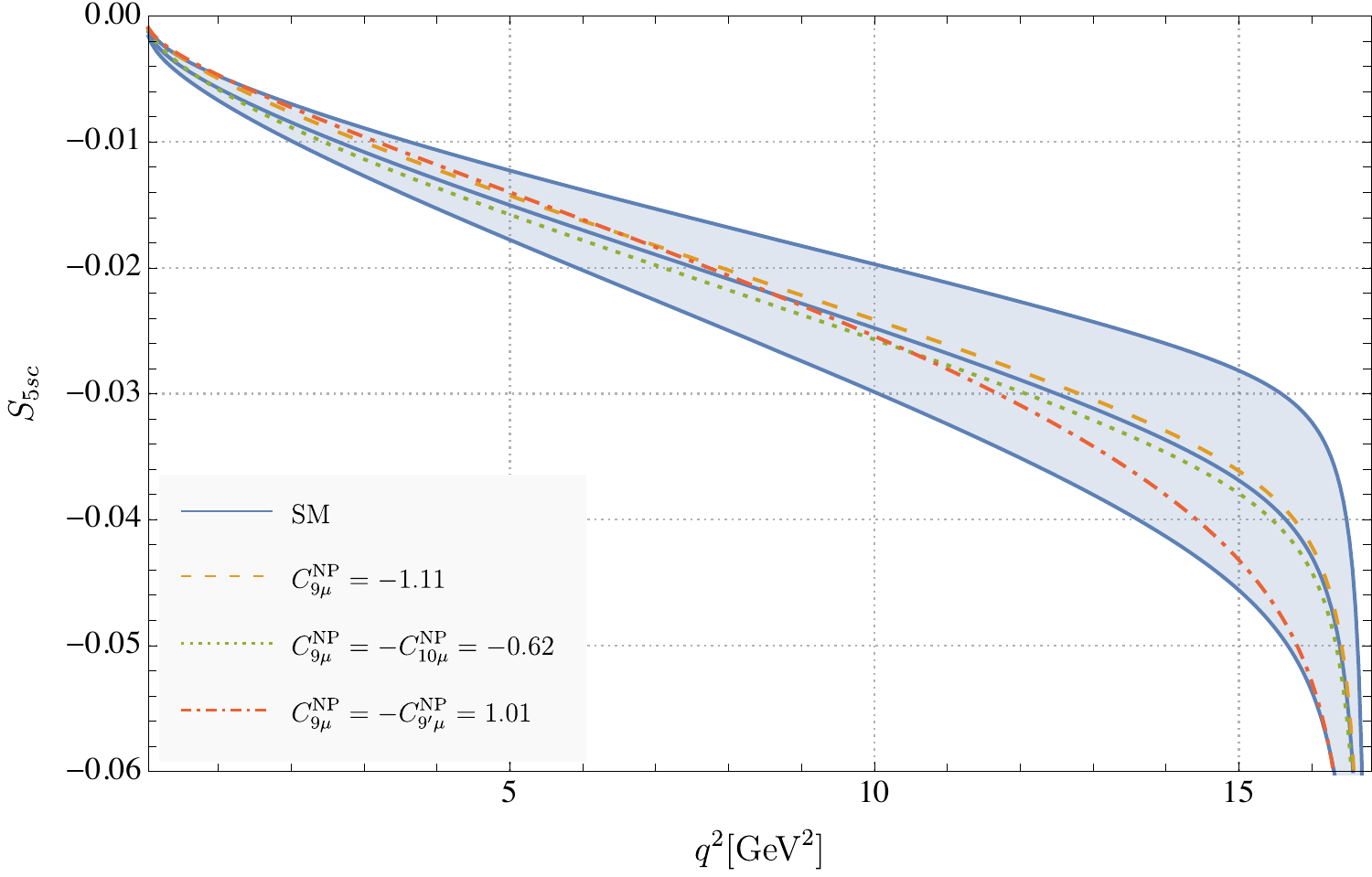}
    \caption{Variation of $S_{3ss}$ (top) and $S_{5sc}$ (bottom) with respect to the dilepton invariant mass, in the case of the SM and three NP models.}
    \label{fig:Sfig2}
\end{figure}

First we show the differential decay rate (normalised to the total $\Lb$ decay width) and the lepton forward-backward asymmetry as functions of the dilepton invariant mass in Figs.~\ref{fig:decayrate} and \ref{fig:AFB} in the context of the SM and several NP models inspired by a recent global fit to $b\to s\ell^+\ell^-$ transitions~\cite{Capdevila:2017bsm}: $\C{9\mu}^{\rm NP}=-1.11$,
$\C{9\mu}^{\rm NP}=-\C{9'\mu}^{\rm NP}=-0.62$,  $\C{9\mu}^{\rm NP}=-\C{10\mu}^{\rm NP}=-0.62$.
We see that the normalisation of the branching ratio is significantly affected by the presence of NP, while keeping a similar shape in all cases. The lepton forward-backward asymmetry exhibits a zero at large recoil whose position depends on the scenario considered. The mild kink at $q^2=4m_c^2$ corresponds to the opening of the $c\bar{c}$ threshold, appearing at the perturbative level as an imaginary part in $\C{9}^{\rm eff}(q^2)$ for $q^2\geq 4m_c^2$.

Due to the lack of accurate estimates of the uncertainties for the various hadronic inputs, we
present only a very crude estimate of the uncertainties, assuming uncorrelated 10\% uncertainties for the $f_{0,\perp,t}$ form factors and uncorrelated 30\% uncertainties for the $f_g$ form factors, on the basis of the expected accuracy from lattice determinations~\cite{Meinel:2016cxo,MeinelPrivate}.
The corresponding error bar for each observable is shown in the case of the SM predictions, but it is
a rather conservative error, as we do not take into account the fact that the various form factors are significantly correlated, as illustrated in both HQET and SCET limits.
In order to keep our figures simple to read, we do not show the uncertainties for the various NP models, which are of the same order as in the SM case.

For illustrative purposes, we also show the impact of a naive model of charmonium resonances in the SM case~\cite{Mott:2011cx}, confirming that the impact of charm loops remains quite small below 8 GeV$^2$ in general, and that the lepton forward-backward asymmetry is much less affected as soon as the resonance region is left.
In Figs.~\ref{fig:decayrate} and \ref{fig:AFB}, the window at low recoil between the $\psi(2S)$ resonance (above 15 GeV$^2$) and the endpoint is rather small, which may affect the application of quark-hadron duality. In the large-recoil region, the main issue is related to non-local contributions from charm loops, which may affect significantly the decay rate but cancels mostly in the lepton forward-backward asymmetry.

We now move to the normalised angular coefficients defined in Eq.~(\ref{eq:SandA}). We do not consider the CP-asymmetries $A$, or the rates $S$ which involve the imaginary part of the products of amplitudes, as these quantities are very dependent on assumptions about the phase of these amplitudes, and in particular the charm-loop contributions. We show the most interesting remaining observables in Figs.~\ref{fig:Sfig1} and \ref{fig:Sfig2}.
We see that these normalised angular coefficients are sensitive to the scenario with right-handed contributions $\C{9\mu}^{\rm NP}=-\C{9'\mu}^{\rm NP}$, but the sensitivity is more limited for scenarios with NP contributions in $\C{9\mu}^{\rm NP}$ only or in $\C{9\mu}^{\rm NP}=-\C{10\mu}^{\rm NP}$.
$S_{1c}$ exhibits some sensitivity to these scenarios, with a $q^2$-dependence
very similar to $A^\ell_{FB}$ (the two quantities are actually identical in both HQET and SCET limits).

As expected from HQET and SCET expectations, the form factors $f_g$ do not contribute much to the amplitudes apart from the vicinity of the low-recoil endpoint. In particular, $B$ is small compared to the amplitudes $A$, which explains that most of the angular coefficients have a very similar behaviour, see Eq.~(\ref{eq:K-HQET}). Moreover, in the SM and in the NP models with no right-handed currents (NP in $\C{9\mu}^{\rm NP}$ only or in $\C{10\mu}^{\rm NP}=-\C{9\mu}^{\rm NP}$), the four dominant amplitudes $A$ are the left-handed ones ($A_{\perp,||}^L$), with contributions all proportional to $\C{9,10,\pm}^L=\C{9\mu}-\C{10\mu}$.
These contributions are all modified in the same proportion in the presence of NP in $\C{9\mu}$ and/or $\C{10\mu}$. The dependence on the Wilson coefficients cancels out between the numerator and the denominator of the normalised angular coefficients $S_i$, which have thus the same $q^2$-dependence for all these scenarios, as can be seen in Figs.~\ref{fig:Sfig1} and \ref{fig:Sfig2}.
This conclusion holds for most of the physical domain, apart from a region at very small $q^2$ where the photon pole is dominant. Subdominant variations related to the interference between the left-handed contribution $\C{9,10,\pm}^L$ and the other amplitudes (photon pole $\C{7}\pm \C{7'}$, right-handed contributions $\C{9,10,\pm}^R$)
can be seen for $S_{1c}$  and $S_{5sc}$ at large recoil.

On the other hand, the scenario with right-handed contributions $\C{9\mu}^{\rm NP}=-\C{9'\mu}^{\rm NP}$ will affect differently $\C{9,10,+}^L$ and
$\C{9,10,+}^R$, which are the dominant contributions in the normalised angular coefficients. It is thus not surprising that the $q^2$-dependence of these coefficients is rather different for this NP scenario with right-handed couplings, as can be seen from the curves in Figs.~\ref{fig:Sfig1} and \ref{fig:Sfig2} that differ significantly from the SM case.

At the low-recoil endpoint for $q^2\to (m_\Lb-m_\Lst)^2$, the situation is slightly different and it depends on the behaviour of the form factors. In this region,
the model of form factors in Ref.~\cite{Mott:2011cx} is less singular than requested from Eqs.~(\ref{eq:endpointbehaviourV}) and (\ref{eq:endpointbehaviourT}).
By inspecting Eqs.~(\ref{eq:HplusHminus1}) (\ref{eq:HplusHminus2}), (\ref{eq:HplusHminus3})
and (\ref{eq:ABampl}), one can see that
only the contributions from $f_g^V$ and $f_g^T$ form factors survive in the various helicity amplitudes, so that
the angular coefficients
$L_{1c}, L_{2c}, L_{4ss}, L_{5s}, L_{6sc}$ vanish. If we neglect the (very small) contribution from the photon pole (i.e. we take $\C{7},\C{7'}\to 0$), we obtain the following results for the other observables at the low-recoil endpoint
\begin{equation}
    S_{1c}\to 0,\qquad
    S_{2cc}-S_{1cc}/4\to 3/8,\qquad
    S_{3ss}\to -1/4 ,\qquad
    S_{5sc}\to -1/2.
\end{equation}
Apart from $S_{1c}$ which vanishes, these values are significantly larger than those obtained over the rest of the physical region. Indeed, as $f_g^V$ and $f_g^T$ are the only non-vanishing contributions, the normalised differential decay rate $d\Gamma/dq^2/N$ (with $N$ defined in Eq.~(\ref{eq:NormalisationN})) becomes smaller by several orders of magnitudes when $q^2\to (m_\Lb-m_\Lst)^2$ and enhances the values of $S_{2cc}-S_{1cc}/4$,
$S_{3ss}$ and $S_{5sc}$ at that endpoint compared to the rest of the physical range for the dilepton invariant mass.

Our study is of course very preliminary and should be refined in several ways in order to provide accurate predictions beyond this exploratory work:
we have no inputs on the form factors determined from lattice simulations or light-cone sum rules,
we have included no correlations among the uncertainties on these form factors even though they are correlated in both SCET and HQET limits. Moreover, we do not attempt any assessment of the charmonium contribution. All these issues should be discussed before drawing definite conclusions concerning the sensitivity of these observables to NP contributions.

\section{Outlook}\label{sec:concl}

We have investigated the rare decay $\Lb\to \Lst(\to N\bar{K})\ell^+\ell^-$ as a new source of information on the flavour-changing neutral-current $b\to s\ell^+\ell^-$ transitions ($\ell=e,\mu$), in addition to the meson channels already studied at $B$ factories and the LHC, which exhibit interesting patterns of deviations compared to the Standard Model (SM) expectations.
We gave a detailed description of the kinematics of the decay and emphasised the issues related to the propagation and the strong decay of the spin-$3/2$ $\Lst$ baryon. We computed the decay rate within the effective Hamiltonian approach, considering only SM and chirality-flipped operators, taking the narrow-width approximation for the $\Lst$ baryon. The resulting differential decay rate is expressed in terms of 12 angular observables that depend on the dilepton invariant mass $q^2$. Each observable can be seen as the sum of interference terms among 12 helicity amplitudes, which can be expressed in terms of short-distance Wilson coefficients and hadronic transition form factors defined in a helicity basis. We checked that our result is in agreement with general expectations from the helicity amplitude formalism, and we also checked that our expressions exhibit the expected behaviour in the real-photon limit $q^2\to 0$ in order to recover the branching ratio for $\Lb\to \Lst\gamma$.
We discussed the simplifications arising in the limit of a heavy $b$-quark: depending on the kinematics (low or large $\Lst$ recoil, i.e., large or small $q^2$), the Heavy Quark Effective Theory and the Soft-Collinear Effective Theory can be used to express all the form factors in terms of 2 or 1 reduced form factors at leading order (i.e up to corrections of order $\alpha_s$ and $\Lambda/m_b$).

As there is currently no determination of the form factors available from lattice simulations or light-cone sum rules, we performed a first illustration of the sensitivity of the observables to New Physics contributions using hadronic inputs from quark models. We considered several NP scenarios favoured by the anomalies observed recently in $b\to s\ell^+\ell^-$ decay modes and we compared the results obtained using the whole set of form factors or exploiting the HQET/SCET relations among the form factors. We discussed the phenomenological consequences for some observables. We noticed that the differential decay rate is quite sensitive to the specific NP scenario considered, both at low and large recoils. On the other hand, the angular coefficients normalised to this decay rate show fewer variations. Indeed, in the case of NP scenarios with moderate contributions to $\C{9\mu}^{\rm NP}$ and/or $\C{10\mu}^{\rm NP}$, the four numerically significant amplitudes ($A_{\perp,||}^L$) are dominated by a single combination of Wilson coefficients which cancel between the numerator and the denominator of the  angular coefficients $S$ normalised with respect to the branching ratio.
In the very large-recoil region, the interference with the photon pole
allows for some discrimination between the NP scenarios for some of the observables. On the other hand, these angular coefficients turn out to be quite sensitive to the presence of right-handed contributions $\C{9'\mu}^{\rm NP}$ which affect differently the various dominant transversity amplitudes. These conclusions are only qualitative: form factors with a better control of theoretical uncertainties should be used to analyse the sensitivities of these observables to NP contributions in more detail before drawing any final conclusions.

Future experimental information on these observables could thus provide complementary information the on-going search for new physics from $b\to s\ell^+\ell^-$ transitions. However, several issues must be solved before this mode can be competitive compared to $B\to K(^*)\ell^+\ell^-$ and even $\Lb\to\Lambda(\to N\pi)\ell^+\ell^-$ decays.
Indeed, the capacity of the LHCb experiment to observe this decay remains to be demonstrated, and the theoretical determination of hadronic contributions, local (form factors) and non-local (charm loops) has to be performed accurately. In principle, one could also exploit the polarisation of the initial and final state to build further observables, similarly to Ref.~\cite{Blake:2017une} in the $\Lb\to\Lambda\ell^+\ell^-$ case.
These aspects should be investigated and solved (partially or fully) in the future. This would
open the possibility for a study of $\Lb\to \Lst(\to N\bar{K})\ell^+\ell^-$ at LHC that could complement other modes in the ongoing quest for New Physics in $b\to s\ell^+\ell^-$ transitions.

\acknowledgments

We would like to thank Y.~Amhis and C.~Marin Benito for very fruitful and enjoyable discussions on the topics covered in this article, V.~Bernard for her insights on the treatment of the $\Delta$ baryon,  J.~Charles, D.~van Dyk and S.~Meinel for useful exchanges concerning the $\Lb\to\Lst$ form factors. We would like also to thank the Rudger Boskovic Institute (Zagreb, Croatia) and the Mainz Institute for Theoretical Physics MITP (Germany) where part of this work was carried out.
This project has received funding from the European Union's Horizon 2020 research and innovation programme under grant agreements No 690575, No 674896, No. 692194 and  No. 692194.

\appendix

\section{Notation}\label{app:kinematics}

\subsection{Kinematics}

In agreement with the general analysis in terms of helicity amplitudes~\cite{Haber:1994pe,Gratrex:2015hna}, we consider the kinematics of the decay in each of the relevant rest frames, which also provides a definition of the angles of interest. In the $\Lb$ rest frame, we have
\begin{equation}
q^\mu=\left(\begin{array}{ccc}\frac{m_\Lb^2+q^2-m_\Lst^2}{2m_\Lb}\\ 0 \\ 0 \\
  -\frac{1}{2m_\Lb}\sqrt{\lambda(m_\Lb^2,m_\Lst^2,q^2)}\end{array}\right),
  \qquad
k^\mu=\left(\begin{array}{ccc}\frac{m_\Lb^2+m_\Lst^2-q^2}{2m_\Lb}\\ 0 \\ 0 \\
  \frac{1}{2m_\Lb}\sqrt{\lambda(m_\Lb^2,m_\Lst^2,q^2)}\end{array}\right)  .
\end{equation}
In the dilepton rest frame (where the basis of polarisation vector $\varepsilon$ is also defined), we have
\begin{equation}
q_1^\mu=\left(\begin{array}{ccc}E_\ell\\ -E_\ell\beta_\ell\sin\theta_\ell \\ 0 \\ -E_\ell\beta_\ell\cos\theta_\ell\end{array}\right),\qquad
q_2^\mu=\left(\begin{array}{ccc}E_\ell\\ E_\ell\beta_\ell\sin\theta_\ell \\ 0 \\ E_\ell\beta_\ell\cos\theta_\ell\end{array}\right),
\end{equation}
where
\begin{equation}
E_\ell=\frac{\sqrt{q^2}}{2}, \qquad \beta_\ell=\sqrt{1-\frac{4m_\ell^2}{q^2}}.
\end{equation}

In the $\Lst$ rest frame we have
\begin{equation}
k_1^\mu=\left(\begin{array}{ccc}\frac{m_\Lst^2+m_N^2-m_{\bar{K}}^2}{2m_\Lst}\\ \frac{m_\Lst}{2}\beta_{N\bar{K}}\sin\theta_\Lst \cos\phi \\
\frac{m_\Lst}{2}\beta_{N\bar{K}}\sin\theta_\Lst \sin\phi \\ \frac{m_\Lst}{2}\beta_{N\bar{K}}\cos\theta_\Lst \end{array}\right),\qquad
k_2^\mu=\left(\begin{array}{ccc}\frac{m_\Lst^2+m_{\bar{K}}^2-m_N^2}{2m_\Lst}\\ -\frac{m_\Lst}{2}\beta_{N\bar{K}}\sin\theta_\Lst \cos\phi \\
-\frac{m_\Lst}{2}\beta_{N\bar{K}}\sin\theta_\Lst \sin\phi \\ -\frac{m_\Lst}{2}\beta_{N\bar{K}}\cos\theta_\Lst \end{array}\right),\end{equation}
where
\begin{equation}
\beta_{N\bar{K}}=\frac{1}{m_\Lst^2}\sqrt{\lambda(m_\Lst^2,m_N^2,m_{\bar{K}}^2)}.
\end{equation}

These definitions agree with the LHCb convention for $\Lb\to\Lambda(\to N\pi)\ell^+\ell^-$~\cite{ Blake:2017une,Aaij:2015xza,Aaij:2018gwm} (up to the identifications $\theta_\Lst=\theta_b$ and $\phi=\chi$) and they also
agree with the LHCb convention for $B\to K^*(\to K\pi)\ell^+\ell^-$ decays~\cite{Aaij:2013iag,Gratrex:2015hna} up to the identification $\{B^0,K^{*0},K^+,\pi^-\}\to    \{\Lb,\Lst,p,K^-\}$.

\subsection{Free solutions in the $\Lb$ rest frame}

For the leptons, we can use the well-known expressions for the spin-1/2 case, see for instance Ref.~\cite{Haber:1994pe} where the application to helicity amplitudes is discussed. We have the following solutions for $\Lb$ for different values for $s_\Lb$
\begin{equation}
u_\Lb(+1/2)=\left(\begin{array}{c} \sqrt{2m_\Lb}\\ 0\\ 0\\ 0 \end{array}\right), \qquad
u_\Lb(-1/2)=\left(\begin{array}{c} 0\\ \sqrt{2m_\Lb}\\ 0\\ 0 \end{array}\right).
\end{equation}

Following Ref.~\cite{Huang:2003ym} as discussed in Sec.~\ref{sec:effhamkin}, we have the solutions for different values for $s_\Lst$
\begin{equation}
\begin{aligned}
u_\Lst(-3/2)=&\frac{1}{2\sqrt{m_\Lb}}\left(\begin{array}{cccc}
0 & 0 & 0 & 0\\
0 & \sqrt{s_+} & 0 & -\sqrt{s_-}\\
0 & -i\sqrt{s_+} & 0 & i\sqrt{s_-}\\
0 & 0 & 0 & 0
\end{array}\right),\\
u_\Lst(-1/2)=&\frac{\sqrt{s_-s_+}}{4\sqrt{3}m_\Lb^{3/2} m_\Lst}\left(\begin{array}{cccc}
0 & 2\sqrt{s_+} & 0 & -2\sqrt{s_-}\\
\frac{2m_\Lst m_\Lb}{\sqrt{s_-}} & 0 & \frac{2m_\Lst m_\Lb}{\sqrt{s_+}} & 0\\
-\frac{2im_\Lst m_\Lb}{\sqrt{s_-}} & 0 & -\frac{2im_\Lst m_\Lb}{\sqrt{s_+}} & 0\\
0 & \frac{s_-+s_+}{\sqrt{s_-}} & 0 &  -\frac{s_-+s_+}{\sqrt{s_+}}
\end{array}\right),\\
u_\Lst(+1/2)=&\frac{\sqrt{s_-s_+}}{4\sqrt{3}m_\Lb^{3/2} m_\Lst}\left(\begin{array}{cccc}
2\sqrt{s_+} & 0 & 2\sqrt{s_-} &0\\
0 & -\frac{2m_\Lst m_\Lb}{\sqrt{s_-}} & 0 & \frac{2m_\Lst m_\Lb}{\sqrt{s_+}} \\
0 & -\frac{2im_\Lst m_\Lb}{\sqrt{s_-}} & 0 & \frac{2im_\Lst m_\Lb}{\sqrt{s_+}} \\
\frac{s_-+s_+}{\sqrt{s_-}} & 0 &  \frac{s_-+s_+}{\sqrt{s_+}} & 0
\end{array}\right),\\
u_\Lst(+3/2)=&\frac{1}{2\sqrt{m_\Lb}}\left(\begin{array}{cccc}
0 & 0 & 0 & 0\\
-\sqrt{s_+} & 0 & -\sqrt{s_-} & 0\\
-i\sqrt{s_+} & 0 & -i\sqrt{s_-} & 0\\
0 & 0 & 0 & 0
\end{array}\right),
\end{aligned}
\end{equation}
where the matrix notation corresponds to the vector and the spinor indices of the solutions $u_{\Lst,a}^\alpha$.

\subsection{Free solutions in the $\Lst$ rest frame}

We have the following solutions for $\Lst$ for different values for $s_\Lst$
\begin{equation}
\begin{aligned}
u_\Lst(-3/2)=&\sqrt{m_\Lst}\left(\begin{array}{cccc}
0 & 0 & 0 & 0\\
0 & 1 & 0 & 0\\
0 & -i & 0 & 0\\
0 & 0 & 0 & 0
\end{array}\right),\qquad
u_\Lst(-1/2)=\sqrt{\frac{m_\Lst}{3}}\left(\begin{array}{cccc}
0 & 0 & 0 & 0\\
1 & 0 & 0 & 0\\
-i & 0 & 0 & 0\\
0 & 2 & 0 & 0
\end{array}\right),\\
u_\Lst(+1/2)=&\sqrt{\frac{m_\Lst}{3}}\left(\begin{array}{cccc}
0 & 0 & 0 & 0\\
0 & -1 & 0 & 0\\
0 & -i & 0 & 0\\
2 & 0 & 0 & 0
\end{array}\right),\qquad
u_\Lst(+3/2)=\sqrt{m_\Lst}\left(\begin{array}{cccc}
0 & 0 & 0 & 0\\
-1 & 0 & 0 & 0\\
-i & 0 & 0 & 0\\
0 & 0 & 0 & 0
\end{array}\right).
\end{aligned}
\end{equation}

We have the following solutions for $N$ for different values for $s_N$
\begin{equation}
u_N(+1/2)=\frac{1}{\sqrt{2m_\Lst}}\left(\begin{array}{c}
\sqrt{r_+}\cos\frac{\theta_\Lst}{2}\\
\sqrt{r_+}\sin\frac{\theta_\Lst}{2}e^{i\phi}\\
\sqrt{r_-}\cos\frac{\theta_\Lst}{2}\\
\sqrt{r_-}\sin\frac{\theta_\Lst}{2}e^{i\phi} \end{array}\right), \qquad
u_N(-1/2)=\frac{1}{\sqrt{2m_\Lst}}\left(\begin{array}{c}
-\sqrt{r_+}\sin\frac{\theta_\Lst}{2}e^{-i\phi}\\
\sqrt{r_+}\cos\frac{\theta_\Lst}{2}\\
\sqrt{r_-}\sin\frac{\theta_\Lst}{2}e^{-i\phi}\\
-\sqrt{r_-}\cos\frac{\theta_\Lst}{2} \end{array}\right) .
\end{equation}

\subsection{Dilepton rest frame}
We have the following solutions for $\ell^-$ for different values for $s_{\ell^-}$
\begin{equation}
u_{\ell^-}(+1/2)=\left(\begin{array}{c}
\sqrt{E_\ell+m_\ell}\cos\frac{\theta_\ell}{2}\\
\sqrt{E_\ell+m_\ell}\sin\frac{\theta_\ell}{2}\\
\sqrt{E_\ell-m_\ell}\cos\frac{\theta_\ell}{2}\\
\sqrt{E_\ell-m_\ell}\sin\frac{\theta_\ell}{2} \end{array}\right), \qquad
u_{\ell^-}(-1/2)=\left(\begin{array}{c}
-\sqrt{E_\ell+m_\ell}\sin\frac{\theta_\ell}{2}\\
\sqrt{E_\ell+m_\ell}\cos\frac{\theta_\ell}{2}\\
\sqrt{E_\ell-m_\ell}\sin\frac{\theta_\ell}{2}\\
-\sqrt{E_\ell-m_\ell}\cos\frac{\theta_\ell}{2} \end{array}\right) ,
\end{equation}
and for $\ell^+$ for different values for $s_{\ell^+}$
\begin{equation}
v_{\ell^+}(+1/2)=\left(\begin{array}{c}
\sqrt{E_\ell-m_\ell}\cos\frac{\theta_\ell}{2}\\
\sqrt{E_\ell-m_\ell}\sin\frac{\theta_\ell}{2}\\
-\sqrt{E_\ell+m_\ell}\cos\frac{\theta_\ell}{2}\\
-\sqrt{E_\ell+m_\ell}\sin\frac{\theta_\ell}{2} \end{array}\right), \qquad
v_{\ell^+}(-1/2)=\left(\begin{array}{c}
\sqrt{E_\ell-m_\ell}\sin\frac{\theta_\ell}{2}\\
-\sqrt{E_\ell-m_\ell}\cos\frac{\theta_\ell}{2}\\
\sqrt{E_\ell+m_\ell}\sin\frac{\theta_\ell}{2}\\
-\sqrt{E_\ell+m_\ell}\cos\frac{\theta_\ell}{2} \end{array}\right) .
\end{equation}

\section{Cross check of the angular decomposition}\label{app:crosscheckang}

The structure of the differential decay rate obtained in Sec.~\ref{sec:angular} can be checked against the general analysis in terms of helicity amplitudes performed in Ref.~\cite{Gratrex:2015hna}. Following the arguments presented there, taking into account the spins of the
initial, intermediate and final states as well as the absence of spin-2 operators in the effective Hamiltonian,
we expect the differential decay rate to be organised as
\begin{equation}\label{eq:partialwaves}
\begin{aligned}
L \propto & {\rm Re}\sum_{L_\Lst=0}^{2J_\Lst} \sum_{L_\ell=0}^{2J_\gamma} \sum_{M=0}^{\min(L_\Lst,L_\ell)}
     G^{L_\Lst,L_\ell}_M(q^2) \Omega^{L_\Lst,L_\ell}_M(\Omega_\Lst,\Omega_\ell)\\
\propto & {\rm Re}[G^{0,0}_0 \Omega^{0,0}_0
     +G^{0,1}_0 \Omega^{0,1}_0+G^{0,2}_0 \Omega^{0,2}_0\\
&\qquad    +G^{2,0}_0 \Omega^{2,0}_0+G^{2,1}_0 \Omega^{2,1}_0
    +G^{2,1}_1 \Omega^{2,1}_1+G^{2,2}_0  \Omega^{2,2}_0+G^{2,2}_1  \Omega^{2,2}_1+G^{2,2}_2  \Omega^{2,2}_2].
\end{aligned}
\end{equation}
The index $L_\Lst$ corresponds to the $N\bar{K}$ system, $L_\ell$ to the dilepton system, and $M$
to the $\phi$-component of both partial waves.
In our case we have $J_\Lst=3/2$ and $J_\gamma=1$, which is the maximal spin of the virtual gauge boson induced by the operators of the effective Hamiltonian in the absence of tensor contributions, as discussed in detail in Ref.~\cite{Gratrex:2015hna}. $G$ are angular coefficients depending on the invariant mass of the dilepton pair. The angular functions are given by the product of Wigner $D$ functions
\begin{equation}
\Omega^{L_\Lst,L_\ell}_M(\Omega_\Lst,\Omega_\ell)=D^{\Lst}_{M,0}(\phi,\theta_\Lst,-\phi) D^{L_\ell}_{M,0}(0,\theta_\ell,0).
\end{equation}
The second helicity index of both Wigner functions in the angular distribution is zero, i.e. the difference of the helicities  of the final-state particles (summed incoherently), and the first index, identical for both Wigner functions, contains the helicities of the internal particles (summed coherently).

Although the sum with respect to $L_\Lst$ in the first line of Eq.~(\ref{eq:partialwaves}) goes from 0 to $2J_\Lst$,  the second line of Eq.~(\ref{eq:partialwaves}) contains only the sum over even values of $L_\Lst$: this is due to the fact that the decay of the $\Lst$ baryon is strong and conserves parity, so that it should be invariant under $\theta_\Lst\to \theta_\Lst+\pi$, which eliminates odd-$L_\Lst$ partial waves~\footnote{Similarly, the decay $B\to K^*(\to K\pi)\ell^+\ell^-$ involves a sum over even values of $J_{K^*}$ in Eq.~(28) of Ref.~\cite{Gratrex:2015hna} since $K^*$ decays strongly, whereas  $\Lb\to \Lambda(\to N\pi)\ell^+\ell^-$ involves a sum over odd and even values of $J_\Lambda$ in Eq.~(E.3) of Ref.~\cite{Gratrex:2015hna}  as $\Lambda(1150)$ decays weakly. This is related to the $P$-conserving or violating nature of the decay of the intermediate hadron, and not to the nature of the final state (as stated in Ref.~\cite{Gratrex:2015hna}).}.
In addition, some of the Wigner functions are real and $L$ is the real part of the product of these functions with the angular coefficients $G$, which means that only the following 12 angular coefficients are involved
\begin{equation}
\begin{aligned}
  {\rm Re}\ G^{0,0}_0\to &\frac{1}{9} (L_{1cc}+2 L_{1ss}+2L_{2cc}+4L_{2ss}+2L_{3ss}),\\
   {\rm Re}\ G^{0,1}_0\to &\frac{1}{3} (L_{1c}+2 L_{2c}),\\
   {\rm Re}\ G^{0,2}_0\to &\frac{2}{9} (L_{1cc}-L_{1ss}+2 L_{2cc}-2 L_{2ss}-L_{3ss}),\\
   {\rm Re}\ G^{2,0}_0\to &\frac{2}{9} (L_{1cc}+2 L_{1ss}-L_{2cc}-2 L_{2ss}-L_{3ss}),\\
   {\rm Re}\ G^{2,1}_0\to &\frac{2 (L_{1c}-L_{2c})}{3},\\
   {\rm Re}\ G^{2,1}_1\to &\frac{2 L_{5s}}{\sqrt{3}},\qquad
   {\rm Im}\ G^{2,1}_1\to -\frac{2 L_{6s}}{\sqrt{3}},\\
   {\rm Re}\ G^{2,2}_0\to &\frac{2}{9} (2 L_{1cc}-2 L_{1ss}-2 L_{2cc}+2 L_{2ss}+L_{3ss}),\\
   {\rm Re}\ G^{2,2}_1\to &\frac{2 L_{5sc}}{3},\qquad
   {\rm Im}\ G^{2,2}_1\to -\frac{2 L_{6sc}}{3},\\
   {\rm Re}\ G^{2,2}_2\to &\frac{4 L_{3ss}}{3},\qquad
   {\rm Im}\ G^{2,2}_2\to -\frac{4 L_{4ss}}{3},
 \end{aligned}
\end{equation}
where we have indicated the equivalence with the angular coefficients defined in Eq.~(\ref{eq:angobs}).

\section{Connection with $\Lb\to \Lst(\to N\bar{K}) \gamma$}\label{app:crosscheckgamma}

\subsection{Tensor form factors}

The expressions for $\Lb\to \Lst(\to N\bar{K}) \ell^+\ell^-$ contain a pole at $q^2=0$, which is related to the decay $\Lb\to \Lst(\to N\bar{K}) \gamma$. The matrix element responsible for the photon contribution to  $\Lb\to\Lst\ell^+\ell^-$ will have the structures $\epsilon^*_\mu M^\mu$ and $\epsilon^*_\mu M^\mu_5$ with
\begin{equation}
M^\mu=\bar{u}_\alpha \Gamma^{\alpha\mu} u,\qquad M^\mu_5=\bar{u}_\alpha\gamma^5 \Gamma^{\alpha\mu}_5 u,
\end{equation}
with the general form factor decomposition
\begin{equation}
    \Gamma^{\alpha\mu}_{(5)}=q^\alpha\gamma^\mu G_1^{(\prime)}
        +q^\alpha (p+k)^\mu G_2^{(\prime)}
        +q^\alpha q^\mu G_3^{(\prime)}
        -g^{\alpha\mu} G_4^{(\prime)}.
\end{equation}
The gauge condition $q_\mu\Gamma^{\alpha\mu}_{(5)}=0$ implies that
\begin{eqnarray}
    G_4&=(m_\Lb-m_\Lst)G_1+(m_\Lb^2-m_\Lst^2)G_2+q^2G_3,\\
    G_4'&=(m_\Lb+m_\Lst)G_1'+(m_\Lb^2-m_\Lst^2)G_2'+q^2G_3',
\end{eqnarray}
leading to the expressions
\begin{equation}
     \Gamma^{\alpha\mu}=[q^\alpha\gamma^\mu-g^{\alpha\mu}(m_\Lb-m_\Lst)] G_1
        +[q^\alpha (p+k)^\mu-g^{\alpha\mu}(m_\Lb^2-m_\Lst^2)]G_2
        +[q^\alpha q^\mu-g^{\alpha\mu}q^2]G_3,
\end{equation}
\begin{equation}
     \Gamma^{\alpha\mu}_5=[q^\alpha\gamma^\mu-g^{\alpha\mu}(m_\Lb+m_\Lst)] G_1'
        +[q^\alpha (p+k)^\mu-g^{\alpha\mu}(m_\Lb^2-m_\Lst^2)]G_2'
        +[q^\alpha q^\mu-g^{\alpha\mu}q^2]G_3'.
\end{equation}

Focusing on the tensor form factors $f_i^T$ needed for the contribution of $\C7+\C{7'}$ given in Eq.~(\ref{eq:ffT}), we can see that we have the identification
\begin{eqnarray}
    G_1 &\to& -(m_\Lb+m_\Lst)f_\perp^T-\frac{m_\Lst}{s_-}f_g^T,\\
    G_2 &\to& -\frac{q^2}{s_+}f_0^T+\frac{(m_\Lb+m_\Lst)^2}{s_+} f_\perp^T+\frac{m_\Lb(m_\Lb+m_\Lst)-q^2}{s_+s_-}f_g^T,\\
    G_3 &\to& \frac{(m_\Lb^2-m_\Lst^2)}{s_+}[f_0^T-f_\perp^T]-\frac{m_\Lb^2+m_\Lb m_\Lst + 2m_\Lst^2-q^2}{s_+s_-}f_g^T.
\end{eqnarray}
Similarly we have the identification for the  pseudotensor form factors $f_i^{T5}$ for $\C7-\C{7'}$
\begin{eqnarray}
    G_1' &\to& -(m_\Lb-m_\Lst)f_\perp^{T5}+\frac{m_\Lst}{s_+}f_g^{T5},\\
    G_2' &\to& -\frac{q^2}{s_-}f_0^{T5}+\frac{(m_\Lb-m_\Lst)^2}{s_-} f_\perp^{T5}+\frac{m_\Lb(m_\Lb-m_\Lst)-q^2}{s_+s_-}f_g^{T5},\\
    G_3' &\to& \frac{(m_\Lb^2-m_\Lst^2)}{s_-}[f_0^{T5}-f_\perp^{T5}]-\frac{m_\Lb^2-m_\Lb m_\Lst+2m_\Lst^2-q^2}{s_+s_-}f_g^{T5}.
\end{eqnarray}
If we want to have three independent form factors $G_1^{(\prime)}$, $G_2^{(\prime)}$, $G_3^{(\prime)}$ with a finite limit at $q^2\to 0$, it is sufficient to request that $f_\perp^{T(5)},f_0^{T(5)},f_g^{T(5)}$ tend all to a finite value in this limit, see Eq.~(\ref{eq:tensorq2zero}).
Let us emphasise that these conditions are obtained by considering solely the behaviour of the amplitude at $q^2\to 0$ in QCD, so that $O(1)$ means $O((q^2)^0)$ here. The SCET limit, though related, is slightly different, taking $q^2=O(\Lambda^2)$, $m_b\to\infty$ and $\Lambda/m_b\to 0$. Eq.~(\ref{eq:ffSCET}) obtained in the SCET limit shows that the condition for $f_g^{T(5)}$ should be understood as $f_0^{T,T5}(q^2)=O(\Lambda^2/m_b^2)$ then.

Moreover, it is possible to determine relations between some of the tensor form factors at $q^2=0$. Indeed,
 the two matrix elements used in Eq.~(\ref{eq:ffT}) can be obtained from the matrix element
$\braket{\Lst | \bar si \sigma^{\mu\nu} b|\Lb}$
thanks to the identity
$\sigma_{\mu\nu}\gamma_5=i\epsilon_{\mu\nu\rho\sigma}\sigma^{\rho\sigma}/2$.
The latter matrix element can be parametrised in terms of six form factors
given in Ref.~\cite{Mott:2011cx}, which can be used to express all the form factors in Eq.~(\ref{eq:ffT}). These relations yield in particular the very simple relationships at $q^2=0$
\begin{equation}
    G_2=G'_2,\qquad G_1=G'_1-2m_\Lst G'_2,
\end{equation}
leading to the following relations between the form factors in Eq.~(\ref{eq:ffT})
\begin{equation}\label{eq:ffTatq20}
    f_\perp^{T5}(0)=f_\perp^{T}(0),
        \qquad
  f_g^{T5}(0)=f_g^{T}(0)    \frac{m_\Lb+m_\Lst}{m_\Lb-m_\Lst},
\end{equation}

\subsection{Branching ratio}

The branching ratio for radiative decay $\Lb\to\Lst\gamma$ is proportional to
\begin{equation}
    \lim_{q^2\to 0} (q^2 \sum_{X=A,B} |X|^2),
\end{equation}
where the sum goes over the 12 transversity amplitudes in Eq.~(\ref{eq:ABampl}).
If we consider the transversity amplitudes of interest~\footnote{The vector/axial form factors are expected to have a finite limit at $q^2=0$ with specific linear combinations of $f_t^V,f_\perp^V$ and $f_t^A,f_\perp^A$ expected to vanish, as indicated in Sec.~\ref{sec:hadronic}.}, we see that we have the
behaviours given in Tab.~\ref{tab:softphoton} for $q^2\to 0$ in the SM.
From Tab.~\ref{tab:softphoton},
we can see that the only contributions
to the radiative decay
comes from the. $\C{7},\C{7'}$ operators for the transitions $\pm 1/2\to \mp 1/2$ and $\pm 1/2\to \pm 3/2$, whereas the branching ratio $\Lb\to\Lst\gamma$ gets no contributions from the transitions $\pm 1/2\to \mp 1/2$. This situation is naturally reminescent of $B\to K^*\gamma$~\cite{Altmannshofer:2008dz} that gets contributions from the amplitudes with transverse polarisations, but not from longitudinal polarisation, as can be seen as the level of the transversity amplitudes ($1/q^2$ pole in $A_{\perp,||}$ but not in $A_0$).

\begin{table}
\begin{equation}
    \begin{array}{c|c|c|c|c}
     s_\Lb & s_\Lst & H^{V,A} & H^{T,T5} & A,B\\
     \hline
    \pm 1/2 & \pm 1/2 & \frac{1}{\sqrt{q^2}} & \sqrt{q^2} & a \C{9\ell} + b \C{10,\ell} + c \C{7}\\
    \pm 1/2 & \mp 1/2 & 1 & 1 & \sqrt{q^2}\left[a' \C{9\ell} + b'\C{10,\ell}+c' \frac{\C{7}}{q^2}\right]\\
    \pm 1/2 & \pm 3/2 & 1 & 1 & \sqrt{q^2}\left[a'' \C{9\ell} + b'' \C{10,\ell}+c'' \frac{\C{7}}{q^2}\right]
    \end{array}
\end{equation}
\caption{Behaviour of the amplitudes for $\Lb\to \Lst(\to N\bar{K})\ell^+\ell^-$ in the $q^2\to 0$ limit in the SM. $a,b$ are generic numbers coming from the kinematics and the form factors.}\label{tab:softphoton}
\end{table}

This pattern is in agreement with the general arguments developed in Refs.~\cite{Hiller:2007ur,Legger:2006cq}
 for $\Lb\to \Lst(\to N\bar{K}) \gamma$.
 We can therefore link our results further with  the expressions in Ref.~\cite{Legger:2006cq}.
The latter are given with respect to an arbitrary quantisation axis, which we have identified with the $z$-axis defined along the $\Lst$ momentum
in the $\Lb$ rest frame (meaning $\theta_\Lst=0$ and $\phi_\Lambda$ arbitrary, to be integrated over, in the notation of Ref.~\cite{Legger:2006cq}) and for an arbitrary $\Lb$ polarisation which we take $P_{\Lb}=0$, leading to a decay rate proportional to:
\begin{equation}
2[|C_{1/2,1}|^2+|C_{-1/2,-1}|^2]\cos^2\theta_\Lst+
\frac{1}{2}[3|C_{3/2,1}|^2+3|C_{-3/2,-1}|^2+|C_{1/2,1}|^2+|C_{-1/2,-1}|^2] \sin^2\theta_\Lst.
\end{equation}
We can make contact with our expressions by integrating Eq.~(\ref{eq:angobs}) over $\theta_\ell$ and $\phi$, leading to
a decay rate proportional to
 \begin{equation}
    (L_{1cc} + 2 L_{1ss}) \cos\theta_\Lst^2 + (L_{2cc} + 2 L_{2ss} + L_{3ss}) \sin\theta_\Lst^2,
\end{equation}
so that it appears that up to a common normalisation we have the identifications
\begin{equation}
\begin{aligned}
|C_{1/2,1}|^2+|C_{-1/2,-1}|^2
\leftrightarrow &
 |A^L_{||0}|^2+|A^L_{\perp 0}|^2+
 |A^L_{||1}|^2+|A^L_{\perp 1}|^2+ (L\leftrightarrow R),\\
|C_{3/2,1}|^2+|C_{-3/2,-1}|^2
\leftrightarrow &
 |B^L_{||1}|^2+|B^L_{\perp 1}|^2+ (L\leftrightarrow R) ,
 \end{aligned}
\end{equation}
in agreement with the definitions of $A$ and $B$ amplitudes that involve $3/2$ and $1/2$ $\Lst$ polarisations respectively.

\subsection{Matching of the form factors}

A final comment is in order concerning the comparison of our formulae with Ref.~\cite{Hiller:2007ur}. There are three form factors contributing to $\epsilon^*_\mu M^\mu$ at $q^2=0$. However the computation in Ref.~\cite{Hiller:2007ur} involve only the values of two form factors at $q^2=0$. Indeed, the computation of the branching ratio amounts to summing over the physical polarisations, leading to the computation of $M^\mu M^*_\mu$. Since the three tensors involved in $M^\mu$ are all transverse with respect to $q^\mu$, one can check that $G_1$ and $G_2$, but not $G_3$, will contribute to the branching ratio $\Lb \to \Lst\gamma$. Comparing the expressions of the matrix elements $\langle\Lst|\bar{s}\sigma_{\mu\nu}q^\nu b|\Lb\rangle$
and $\langle\Lst|\bar{s}\sigma_{\mu\nu}\gamma_5 q^\nu b|\Lb\rangle$
in Eq.~(\ref{eq:ffT}) and in Ref.~\cite{Hiller:2007ur}
at $q^2=0$,
we obtain the relationships

\begin{eqnarray}
    i\frac{f_1}{2m_\Lb}&=& - f_\perp^T(0) - f_g^T(0)\frac{m_\Lb}{(m_\Lb-m_\Lst)(m_\Lb^2-m_\Lst^2)}\\\nonumber &=& - f_\perp^{T5}(0)- f_g^{T5}(0)\frac{m_\Lb}{(m_\Lb+m_\Lst)(m_\Lb^2-m_\Lst^2)},\\
    if_2&=&(m_\Lb+m_\Lst)f_\perp^T(0)+
    f_g^T(0)\frac{m_\Lst}{(m_\Lb-m_\Lst)^2}\\\nonumber
    &=&(m_\Lb+m_\Lst)f_\perp^{T5}(0)+
    f_g^{T5}(0)\frac{m_\Lst}{m_\Lb^2-m_\Lst^2}
\end{eqnarray}
which agree with the constraints in Eq.~(\ref{eq:ffTatq20}).
We thus check that only two constants arise for $\Lb\to pK^-\gamma$, as proposed in
Ref.~\cite{Hiller:2007ur}.
The identification of $f_1,f_2$ with the values of the tensor form factors actually yields further cross-checks with this reference in the SCET limit. Using the relations in Eq.~(\ref{eq:ffSCET}), we see that the contribution proportional to $\tilde{f}_0^T(0)$ can then be neglected, leading to the relation
$f_1=-2f_2m_\Lb/(m_\Lb+m_\Lst)$ given in
Ref.~\cite{Hiller:2007ur}. As discussed in this reference, in the same SCET limit, the amplitudes $C_{\pm 3/2,\pm 1/2}$ indeed vanish, since they correspond to $B_{\perp 1}$ and $B_{||1}$, proportional to $f_g$ and $f^T_g$.

\end{document}